\documentclass[12pt]{article}
\usepackage{epsfig}
\usepackage{bbm} 
\usepackage{bm}

\usepackage{exscale}

\font\fourteenbf=cmbx12 scaled\magstep1

\newcommand{\lsi}{\raise0.3ex\hbox{$<$\kern-0.75em\raise-1.1ex\hbox{$\sim$}}}
\newcommand{\gsi}{\raise0.3ex\hbox{$>$\kern-0.75em\raise-1.1ex\hbox{$\sim$}}}
\newcommand{\lsim}{\mathop{\lsi}}
\newcommand{\gsim}{\mathop{\gsi}}
\renewcommand{\vec}[1]{{\bm #1}}
\newcommand{\lasym}{L_{\rm asym}}
\newcommand{\lmax}{L_{\rm max}}

\begin{document}  
\setlength{\baselineskip}{0.6cm}
\newcommand{\figysize}{16.0cm}
\newcommand{\figtopspace}{\vspace*{-1.5cm}}
\newcommand{\figbottomspace}{\vspace*{-5.0cm}}

\begin{titlepage}
\begin{centering}
\vfill

{\fourteenbf \centerline{Non-abelian plasma instabilities for strong anisotropy
                        }
   \vskip 2mm \centerline{}
}

\vspace{1cm}

Dietrich B\"odeker $ {}^ {\rm a} $\footnote{bodeker@physik.uni-bielefeld.de} 
and  Kari Rummukainen $ {}^ {\rm b} $\footnote{kari.rummukainen@oulu.fi}

\vspace{.6cm} { \em
$ {}^ {\rm a} $  Fakult\"at f\"ur Physik, Universit\"at Bielefeld, D-33615 Bielefeld, Germany 
\\}

\vspace{0.3cm}

{\em $ {} ^{\rm b}$ Department of Physics, University of Oulu, P.O.Box 3000, FI-90014 Oulu, Finland
} 

\vspace{0.3cm}                    


\vspace{2cm}   
 
{\bf Abstract} 

\end{centering}    
 
\vspace{0.5cm}

\noindent 
We numerically investigate gauge field instabilities in 
anisotropic SU(2) plasmas using weak field initial conditions.
The growth of unstable modes is
stopped by non-abelian effects for moderate anisotropy. 
If we increase the anisotropy the growth continues beyond the non-abelian
saturation bound. 
We find strong indications that the
continued growth is not due to over-saturation of infrared field modes,
but instead due to very rapid growth of high
momentum modes which are not unstable in the weak field
limit. 
The saturation amplitude strongly depends on the initial conditions.
For strong initial fields we do not observe the sustained growth. 

\vspace{0.5cm}\noindent

 
\vspace{0.3cm}\noindent
 
\vfill \vfill
\noindent
 
\end{titlepage}

\section{Introduction}

It is  an interesting and still open question 
to what degree the medium created in high energy heavy ion collisions 
reaches local thermal equilibrium before it falls apart.
Results from the RHIC experiments
are hinting towards fast thermalization \cite{early}. For sufficiently large
collision energy the relevant running coupling is small and this problem can be addressed
theoretically in a controlled way using perturbative QCD. Even if a full
analytic calculation is not 
possible one should at least be able to obtain parametric estimates for the
thermalization time and the achieved temperature. Remarkably, the solution to
this problem has not been found yet.

Due to the non-isotropic  
expansion the momentum distribution of the produced partons
becomes anisotropic\footnote{For a nice illustration 
see Fig.~1 of Ref.~\cite{arnoldCascade}.}. If the expansion is
mostly 1-dimensional 
along the collision axis, 
the typical longitudinal momenta
become  much smaller than the transverse momenta. 
Anisotropic momentum distributions cause so called plasma\footnote{Here
  ``plasma'' refers to a system of quarks and 
  gluons  
which is not necessarily in thermal equilibrium, while sometimes the term 
``quark-gluon-plasma'' is reserved for  thermalized or almost thermalized
systems.}
instabilities, {\em  i.e.}, 
certain long wave gauge field modes 
grow exponentially so long as their amplitudes are sufficiently small. This is a
collective phenomenon
which is not visible in the kinetic equation approach
used in \cite{wong,muellerBoltzmann,raju,bottomUp}. 
It has been argued that this  effect, which is well known in 
plasma physics,  
will speed up equilibration in heavy ion collisions since the unstable modes 
tend to make the momentum distributions more isotropic \cite{mrowczynskiInitial}.   

There are important  qualitative differences between QED and 
QCD plasma instabilities \cite{arnoldInstabilities}. 
In both cases the growth of  unstable modes is stopped by non-linear effects. In
QED this happens when the amplitude of the unstable modes has become so large
that they deflect a particle momentum by a large angle within 
a distance of one wavelength. This corresponds to gauge field amplitudes 
$ A $ of order $ p / e $ where $ p $ is a typical particle momentum,
henceforth called ''hard''. When the fields 
become this large they have a dramatic effect on the plasma particles since
they instantaneously make the momentum distribution isotropic. 
In QCD the gauge fields are  self-interacting, and the linear approximation
already breaks down at much 
smaller amplitudes $ A  \sim  k/g $ where  
$k \ll p$  
is a characteristic wave vector of an unstable gauge field mode. A crucial question is  whether
these non-linearities stop the 
growth of instabilities. In Ref.~\cite{arnoldAbelianization} it was suggested 
that gluon self-interactions may not saturate the  
instabilities because the system can ``abelianize''  so that 
the unstable modes can grow until they hit the abelian saturation
bound $ A \lsim p/g $. The distribution of 
hard gluons would then quickly become isotropic, and it has been argued 
\cite{arnoldThermalization} that this is  
sufficient for a hydrodynamic description to be applicable even if there is no
local thermal equilibrium. 

The question how plasma instabilities in QCD saturate is thus an important
one. It can be addressed most cleanly by neglecting both the expansion of the
system and the back reaction on the particle momenta. This is sensible because
in the weak coupling limit\footnote{More precisely, one has to consider not
  only weak   gauge coupling but also sufficiently large times where the
  system is sufficiently dilute so that the very notion of particles is
  applicable. In this regime the 
  expansion rate is parametrically small compared to the time scale relevant
  to the instabilities \cite{bottomUp,arnoldInstabilities}.}
the expansion is slow compared to the dynamics of
the unstable modes and because there is a large scale separation of particle
momenta $ p $ and the wave-vectors of unstable modes $ k $.

Because the amplitudes of the unstable field modes become large,
we are dealing with a
non-linear problem and we cannot compute
their time evolution perturbatively. 
So far our qualitative understanding is very limited and one has to
rely on lattice simulations. These are possible due to the large occupation
numbers which allows one to use the classical field approximation for the
infrared fields. 
In lattice simulations with fields depending only on $ t $ and $ z $
it was indeed observed \cite{romatschke2d} that the fields continue to
grow rapidly in the non-linear regime. However, 
3+1 dimensional simulations \cite{arnoldFate,romatschkeFate}
indicate that the instabilities  are saturated  
by non-abelian interactions which would mean  that their  effect is  less
dramatic than suggested in Ref.~\cite{arnoldThermalization}\footnote{For a
  recent discussion of the role of dimensionality see \cite{arnoldLessons}.}. 
In~\cite{impact} it was shown 
that even then the thermalization process {\em  is}  affected by plasma
instabilities, because the broadening of longitudinal momenta
of the particles caused by the unstable modes is more efficient than due to elastic
scattering~\cite{bottomUp}.  

Most lattice simulations have so far been restricted to moderate
anisotropies. 
In the present article we report on the
evolution of instabilities in strongly anisotropic systems. 
In Sec.~\ref{sc:setup} we describe the equations and the approximations we use to
solve them. The results are discussed in Sec.~\ref{sc:results}.
In Ref.~\cite{romatschkeVenugopalan} strongly anisotropic
plasmas have been considered in  a kinematics and with approximations which
are quite different from ours.

\section{The setup}
\label{sc:setup}

Our starting point is the non-abelian Vlasov equations
\cite{heinzKinetic,stanKinetic}   
\begin{eqnarray}
  ( D _  \mu    F ^  {\mu  \nu  } ) ^ a 
  = g \int \frac{ d ^ 3 p }{(2 \pi ) ^   3} v ^ \nu     f ^ a,
  \label{p90} 
\\
  ( v \cdot D f ) ^ a + g v ^ \mu  F _{\mu  i } ^ a 
   \frac{ \partial \bar{f} }{\partial 
    p ^ i} = 0
  \label{vlasov} 
\end{eqnarray}

These are classical equations of motion for SU(2) gauge fields $ A ^ a _ \mu
( x ) $ interacting with particle degrees of freedom.
The average distribution of the particles
$ \bar f  ( \vec p ) \ge 0 $ is a gauge singlet, and the 
leading charged particle density fluctuations are described
by adjoint representation distribution functions $ f ^ a ( x, \vec p )$.  
The particles are moving with the speed of light, thus,
the 3-velocity is $\vec v = \vec p/|\vec p|$, and $ ( v ^ \mu  ) $ is defined
as $  ( 1, \vec v )
$.

We neglect the back reaction
of the soft gauge field $ A _ \mu  $ on $ \bar f   $ and also the
expansion, so we take $ \bar f  ( \vec p ) $ to be space and time independent. 
Neglecting the $ x ^ \mu  $-dependence of $ \bar f $ is justified as long 
as the expansion rate of the system is small compared to the growth rate of the
unstable  
modes we are interested in. 
In an isotropic plasma $  \bar f $ only
depends on $  |  \vec p  |  $; here we consider the anisotropic case, 
but we assume
that $ \bar f $ is invariant when $ \vec p $ is  reflected or rotated around 
the $ z $-axis.

Our equations describe
high momentum modes which are treated as classical colored particles and soft
gluons which are treated as classical fields.  
In order for the classical particle approximation to be valid the wave
 vectors of the fields have to be much smaller than the momenta of the
particles.  The classical field approximation is valid because
we will be dealing with large occupation number (large amplitude)
gluon fields. 
The expansion of the system has been neglected
because at weak coupling the expansion rate is much smaller than the rate at
which the soft gluons evolve. Furthermore, the back-reaction of the soft
fields on the momentum 
distribution has been neglected here ('hard loop approximation').

The $  |  \vec p  |  $-dependence of $ f ^ a $ is irrelevant for determining
the gluon field dynamics. One only needs the integral
\begin{eqnarray}
  W ^ a ( x, \vec v ) \equiv 4 \pi  g \int\limits _ 0 ^ {\infty  } 
  \frac{  d  p  p ^ 2}{(2 \pi )^ 3} f ^ a ( x, p \vec v )  
\end{eqnarray} 
Integrating (\ref{vlasov}) over $  |  \vec p  | $ we obtain
\begin{eqnarray}
  \label{maxwellW}
  ( D _  \mu    F ^ {\mu  \nu  } ) ^ a = \int \frac{ d \Omega  _ \vec v}{4 \pi }
  v ^ \nu  W ^ a 
\\
  \label{vlasovW}
  ( v \cdot D W ) ^ a =  v ^ \mu  F _{\mu  i } ^ a  u ^ i  
\end{eqnarray}
with 
\begin{eqnarray}
  u ^ i (\vec v ) = -4 \pi  g ^ 2  \int\limits _ 0 ^ {\infty  } 
  \frac{  d  p  p ^ 2}{(2 \pi ) ^{3}}  \frac{ \partial \bar{f } ( p \vec v ) }{\partial 
    p ^ i}
\end{eqnarray} 
For  isotropic $ \bar f  $ one would have $ \vec u = m^2_{\rm D} \vec v $, and 
(\ref{vlasov})
would the usual hard thermal loop equation
of motion. For an  anisotropic plasma $ \vec u $ will
not simply be proportional to $ \vec v $. Since we assume  $ \bar f $ to be
parity even,  $ \vec u $ is parity odd. 

As in \cite{lmax} we expand $ W ( x, \vec v ) $ in spherical harmonics, 
\begin{eqnarray}
   W ( x, \vec v ) = \sum _{l = 0} ^{\lmax} \sum _{m = -l} ^{l}  W _ { lm } (
   x ) Y _ { lm } ( \vec v ) 
\end{eqnarray} 
with a finite $ l $-cutoff $ \lmax $. This turns Eqs.~(\ref{maxwellW}),
(\ref{vlasovW}) into classical equations for fields living in $ 3+1 $
dimensions. Similarly we expand 
$ \bar{f} $ in spherical harmonics and we assume that it only depends on 
$ \vec p ^ 2 $ and $ p _ z ^ 2 $. Then
\begin{eqnarray}
  \bar{f} ( \vec p ) = \sum _{l = 0} ^{\lasym}  
  \bar{f} _ l ( |\vec p| ) Y _{l, 0} ( \vec v )
\end{eqnarray}
where the sum runs over even $ l $ only.  In general
the $l$-cutoff $\lasym $ would be infinite, but in practice we must choose
parametrizations with finite $\lasym$ since the equations of motion limit
$\lasym \le \lmax$.  When we increase $\lasym$ it becomes possible to
describe more anisotropic distributions, but at the same time 
$\lmax$ and correspondingly memory- and cpu-time requirements of the
simulations are increased (roughly proportionally to $\lmax^2$).

The equations of motion in terms of $ W _{l m} $ in temporal gauge $ A _ 0 = 0
$ become
\begin{eqnarray}
  \partial _ 0 W _{l m}  + C ^ i _{lm, l'm'} D ^ i W _{l'm'} 
  &=&  
   F _{0i} u ^ i _{lm}   
  +2  F _{i z} u ^{iz} _{lm} 
   \label{eom}  \\
  \partial_0 F^{0i} + D_k F^{ki} &=& v^i_m W_{1m}
   .
   \label{gauge-eom}
\end{eqnarray}
Gauss law reads
\begin{equation}
  D_i F^{i0} = \frac1{\sqrt{4\pi}} W_{00}.  \label{gauss}
\end{equation}
Here $E^i = -F^{0i}$ is the canonical momentum of the gauge field
$A^i$.

The coefficients $ C ^ i _{lm, l'm'} $ may be found in Appendix A of
Ref.~\cite{lmax}. The other coefficients are
\begin{eqnarray} 
   v^i_m = \int \frac{ d\Omega}{4\pi} Y_{1m} v^i, \qquad
   u ^ i _{lm} = \int d \Omega  Y^* _{lm} u ^ i, \qquad 
   u ^ {ij} _{lm} = \frac12 \int d \Omega  Y^* _{lm} (  v ^ i u ^ j - v ^ j u ^ i ) .
\end{eqnarray} 
We now define
\begin{eqnarray}
  m ^ 2 _ l \equiv 4 \sqrt{ \pi  } 
  g ^ 2 
   \int\limits _ 0 ^ {\infty  } 
  \frac{  d  p  p }{(2 \pi ) ^{3}}
  \bar f _ l ( p ) 
  \label{s99.7} 
\end{eqnarray} 
For an isotropic system $ m ^ 2 _ 0 $ equals the Debye mass squared. 
We want $ \bar f $ to be positive which 
gives the condition $ \sum _ l m _ l ^ 2 Y _{l0} ( \vec v ) \ge  0 $
(Albeit we shall violate this condition slightly.).

The only non-vanishing $ u $-coefficients in Eq.~(\ref{eom}) are
\begin{eqnarray}
  \label{s99.8.2.2} 
  u ^ x _{l 1} = - \frac{ \sqrt{\pi  } } { 2 }
  \frac{ \sqrt{l ( l + 1 )} }{\sqrt{2 l + 1}}
    \left ( \frac{ l + 1}{\sqrt{ 2 l - 1}} m ^ 2 _ {l -1}
    + \frac{ l }{\sqrt{2 l + 3}} m ^ 2 _ {l + 1}
  \right ) 
\end{eqnarray}  
\begin{eqnarray}
  u ^ x _{l, - 1} = -  u ^ x _{l 1}, \qquad 
  u ^ y _{l 1} =  u ^ y _{l, - 1} =  -i u ^ x _{l 1}
\end{eqnarray} 
\begin{eqnarray}
  u ^ z _{l 0} = \sqrt{\pi } 
 \frac{ l ( l + 1 ) }{\sqrt{2 l + 1}}
  \left ( \frac{m ^ 2 _ {l -1} }{\sqrt{2 l - 1}} 
    - \frac{ m ^ 2 _ {l + 1}}{\sqrt{2 l + 3}} 
  \right ) 
  \label{s99.6.1.1} 
\end{eqnarray}  

\begin{eqnarray}
  u ^ {xz} _{l 1 } = \frac{ \sqrt{\pi  }}  { 4 } 
  \sqrt{l ( l + 1 )}
 m _ l ^ 2
\end{eqnarray} 

\begin{eqnarray}
  u ^ {xz} _{l, - 1} = -  u ^ {xz} _{l 1}, \qquad 
  u ^ {yz} _{l 1} =  u ^ {yz} _{l, - 1} =  -i u ^ {xz}_{l 1}
\end{eqnarray}

\begin{figure}[t]

  \vspace*{0.5cm}

  \centerline{
              \epsfxsize=8cm
              \epsffile{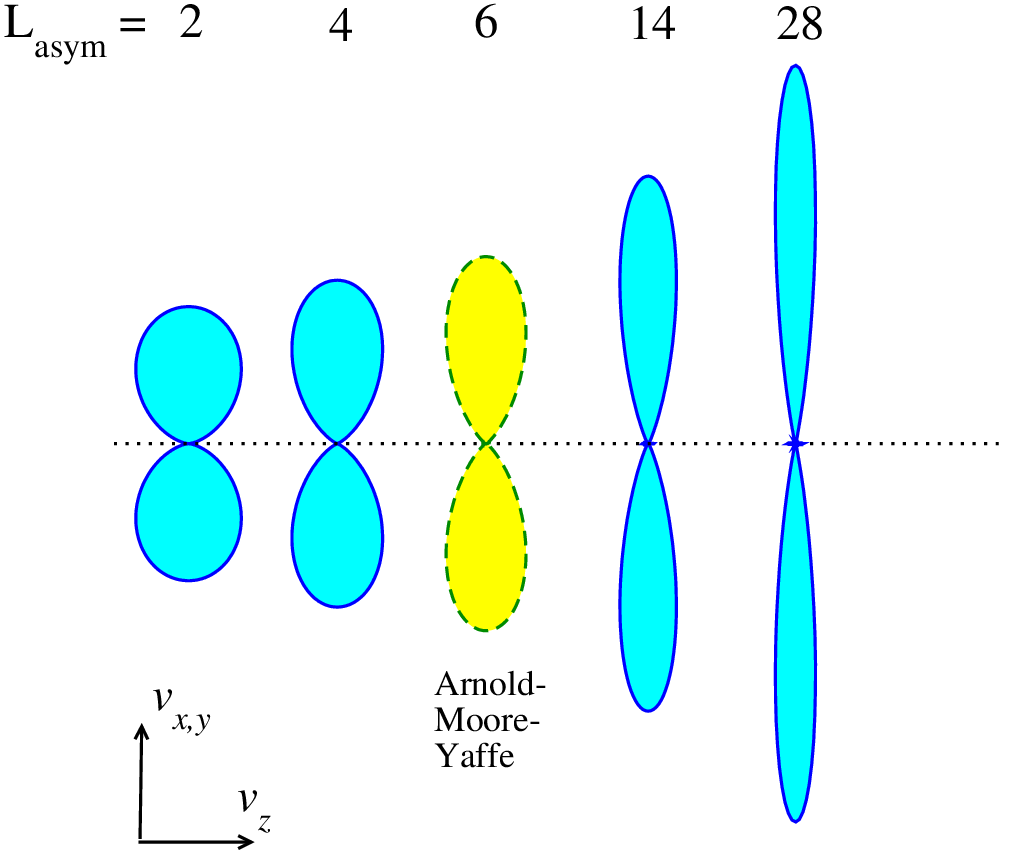}
            }

            \caption{Anisotropic hard particle distributions used in
              this work, together with the distribution used by
              Arnold, Moore and Yaffe \cite{arnoldFate}.  The
              distributions are plotted so that the relative number of
              particles moving to direction $\vec v$ is proportional
              to the length of the radial vector from the center of the
              plot.  For each $
              \lasym $ we tried to maximally localize the distribution
              in the $ x y $-plane.  The distributions are normalized
              to equal area for readability.}
              
  \label{asym_polar}
\end{figure}

We study the behavior of the system using both weakly and strongly 
anisotropic distributions. A measure of the anisotropy is
\begin{equation}
  \eta^2 \equiv 3 \langle v_z^2 \rangle / \langle \vec v^2 \rangle\,,
  \label{eta}
\end{equation}
which equals 1 for symmetric and 0 for completely planar
distribution.

For each $\lasym$ the distribution is parameterized by the coefficients
$m^2_l$, with $l = 0, 2, \ldots, \lasym$.  The values of $m^2_l$ are
chosen so that the anisotropy of the resulting distribution is
approximately maximized.  The reason for this choice is that
for a given anisotropy, we take $\lasym$ as small as possible, also
minimizing the required $\lmax$ and hence computational requirements.

For $\lasym=2$ and 4 the tuning of the
parameters is easy enough to do by hand, but for $\lasym=14$ and 28 we
use a 1-parameter fitting procedure: $\bar f(\theta)$ is fitted to a
narrow Gaussian function centered at $\theta=\pi/2$.  The width of the
Gaussian is adjusted to be as small as possible while still giving a
good fit; if the width of the Gaussian is too small 
the fitted function will have large oscillations over
whole $\theta$-range.  The quality of the fit is justified by eye.
This procedure is sufficient for our purposes: the goal is
to find one good enough parametrization for the asymmetry, and no
attempt is made to maximize the asymmetry for any given
$\lasym$.  The resulting parameters are given in table~\ref{tab:params}.

\begin{table}
\begin{center}
\begin{tabular}{l|cccc}
$\lasym$ &     2    & 4   & 14    &  28 \\
\hline
$\eta^2$  &     0.6  & 0.4  & ~0.086   & ~0.022  \\
\hline
$ m_2^2/m_0^2 $ & -0.447& -0.671 &   -1.021& -1.093  \\
$ m_4^2/m_0^2 $ &       & ~0.167 &   ~0.833& ~1.046  \\
$ m_6^2/m_0^2 $ &       &        &   -0.603& -0.967  \\
$ m_8^2/m_0^2 $ &       &        &   ~0.390& ~0.867 \\
$ m_{10}^2/m_0^2 $ &       &       & -0.227& -0.756  \\
$ m_{12}^2/m_0^2 $ &       &       & ~0.119& ~0.640 \\
$ m_{14}^2/m_0^2 $ &       &       & -0.057& -0.526  \\
$ m_{16}^2/m_0^2 $ &       &       &       & ~0.421 \\
$ m_{18}^2/m_0^2 $ &       &       &       & -0.327  \\
$ m_{20}^2/m_0^2 $ &       &       &       & ~0.247 \\
$ m_{22}^2/m_0^2 $ &       &       &       & -0.181  \\
$ m_{24}^2/m_0^2 $ &       &       &       & ~0.130  \\
$ m_{26}^2/m_0^2 $ &       &       &       & -0.090  \\
$ m_{28}^2/m_0^2 $ &       &       &       & ~0.061 \\
\hline
\end{tabular}
\end{center}
\caption[a]{The parameters $m_l^2$ used in simulations.
$\lasym=2$ and 4 correspond to weak asymmetry, $\lasym=14$ and
28 to strong asymmetry.}
\label{tab:params}
\end{table} 


This process gives distributions where the power is strongly
concentrated around $\theta=\pi/2 \pm \Delta\theta$, where
$\Delta\theta$ is the maximum resolution power of the
$Y_{l0}$-expansion when $l\le\lasym$, that is $\Delta\theta \sim
\pi/\lasym$.  Thus, when plotted on cartesian coordinates, the
distribution has well-defined ``lobes'' centered around direction
$\theta=\pi/2$, i.e. along the $xy$-plane, as shown in
Fig.~\ref{asym_polar}.  For directions near $\theta \approx 0$ or
$\pi$, the distributions can become slightly negative; however, the
magnitude of this effect is negligible.

\begin{figure}[t]

  \vspace*{0.5cm}

  \centerline{
              \epsfxsize=8cm
              \epsffile{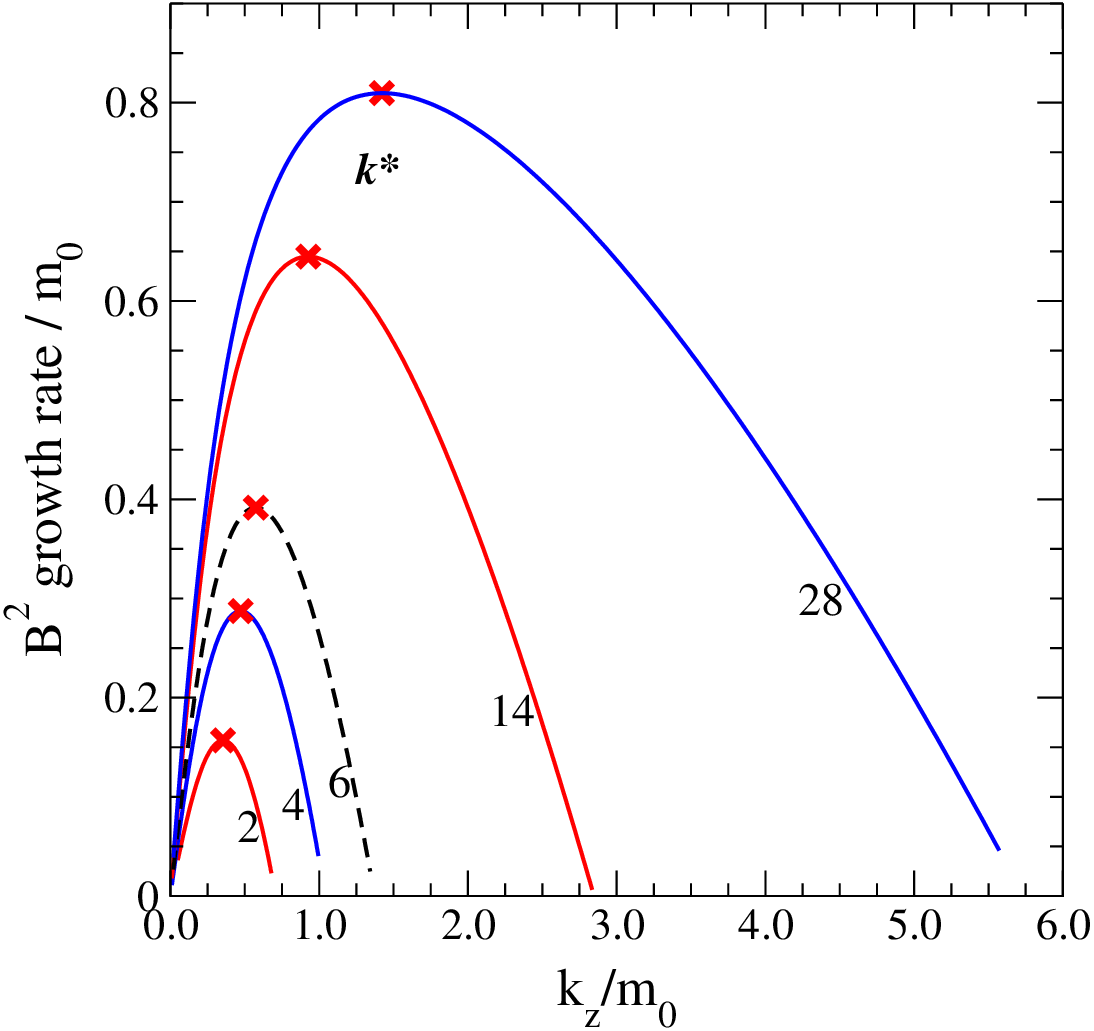}
             }

  \caption{Soft field growth rate 
    as a function of momentum $ \vec k = k \hat { \vec z }
    $ for linearized equations of motion, for 
    anisotropic hard mode distributions $\lasym = 2$, $4$, $6$,
    $14$ and $28$ (see Fig.~\ref{asym_polar} and table~\ref{tab:params}).
    $k_\ast$ is the wave number with the maximal growth rate.
  }
  \label{rate_Lb}
\end{figure}

For small amplitudes the non-linear terms in the equations of motion 
can be neglected.  Modes with different wave vectors do not mix,
and the unstable modes grow exponentially at a rate which can be calculated
analytically. 
The growth rate is shown in Fig.~\ref{rate_Lb} as a function of the
length of the wave vector 
of the unstable mode for different asymmetries. 
For each asymmetry $ k _ \ast $
denotes the value of $ | \vec k | $ for which the growth rate is
maximal.
For the smallest to the largest anisotropy, the maximum growth
rate increases
by a factor of 5 and the width of the unstable mode distribution
by a factor of 8.

\begin{figure}[t]

  \vspace*{0.5cm}
  
  \centerline{
    \epsfysize=7cm
    \epsffile{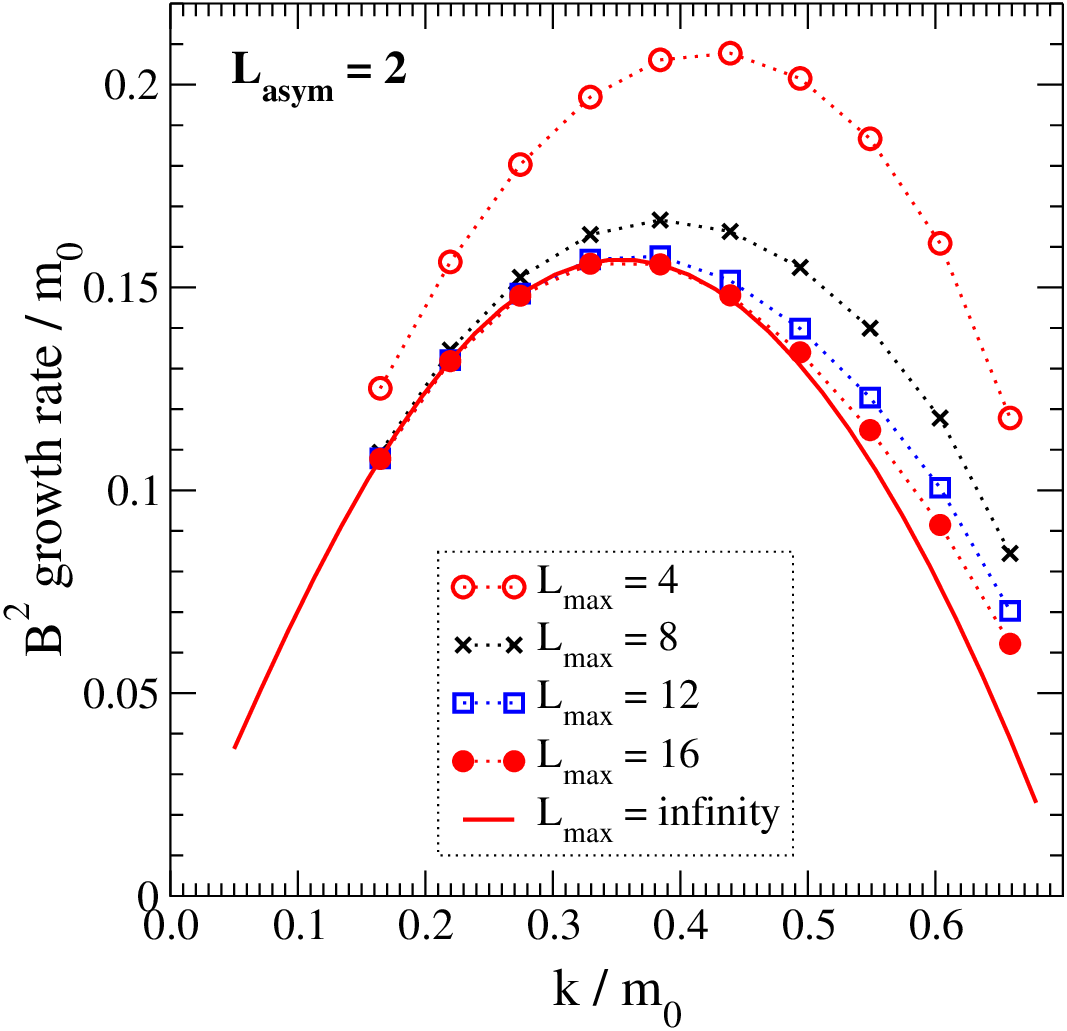}
    \epsfysize=7cm
    \epsffile{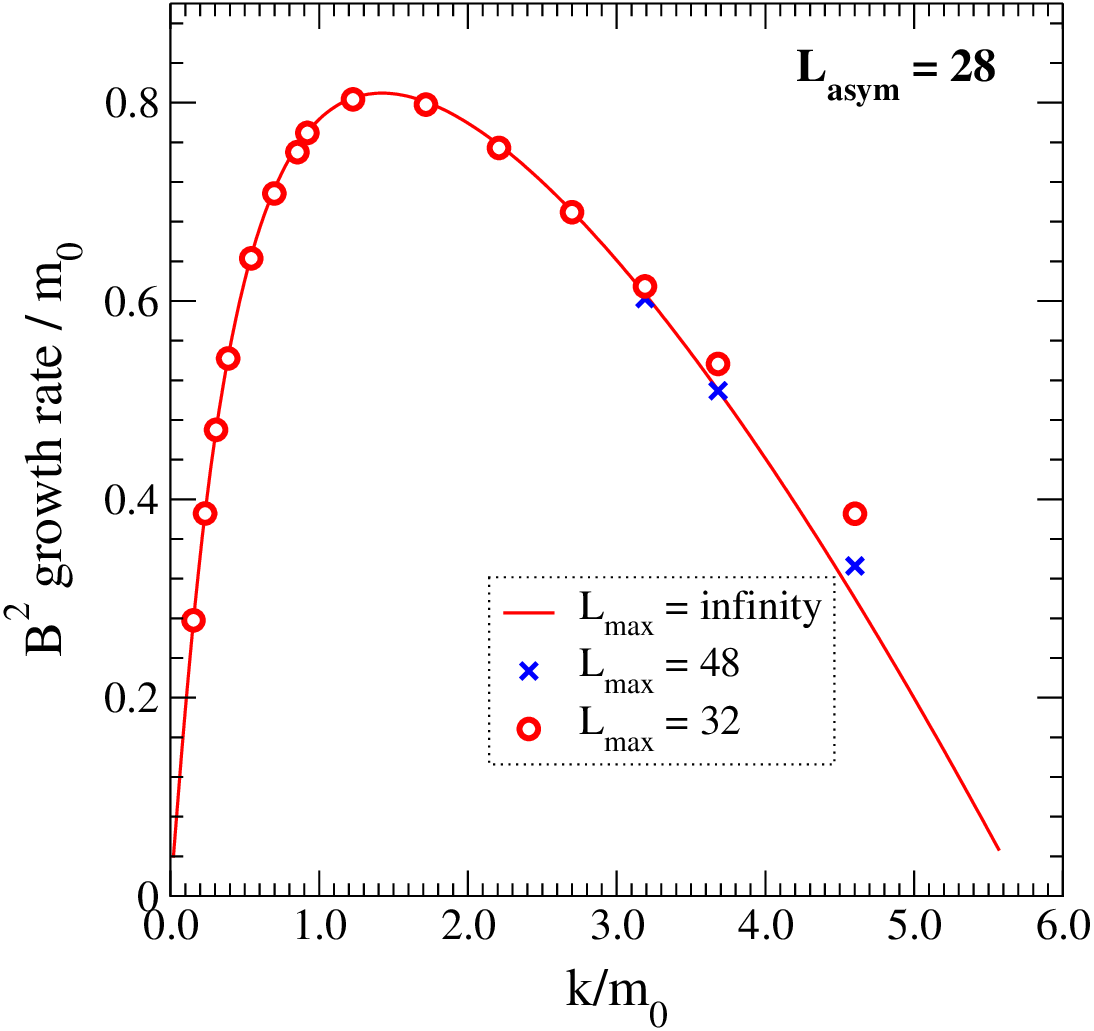}
  }
  \caption{Growth rate of magnetic energy for the linearized equations of
    motion with different $\lmax$ cutoffs, shown for $ \lasym = 2 $  
    (left) and  $
    \lasym = 28 $ (right).  
  }
  \label{rate_L2}
\end{figure}

The linear equations of motion offer a straightforward method for
investigating
how large we need need to make $ \lmax $ in order to reproduce the 
continuum dynamics.   In Fig.~\ref{rate_L2} we compare the
growth rate at $\lmax=\infty$ with the rates at different finite values of $
\lmax $ 
for modes with $ \vec k = k \hat \vec z $.
For weak anisotropy ($ \lasym = 2 $, left figure) one needs rather large values of $ \lmax \gg
\lasym $ to reproduce the 
growth rate. The growth rate for strong anisotropy $ \lasym
= 28 $ (right figure) can be
reproduced already with $ \lmax \gsim  \lasym $.  Indeed, for
the asymmetries used in this study it appears that the
finite $\lmax$ effects are roughly independent of the $\lasym$ used, and we
should obtain accurate results for $\lmax\gsim 16$, of course provided that
we keep $\lmax > \lasym$.  In Sec.~\ref{sc:lattice} we investigate
the $\lmax$-dependence of the real simulations in detail.

We note that while the rate can be solved analytically at finite
$\lmax$, in Fig.~\ref{rate_L2} we actually measured the rate from
numerical simulations using a linearized version of our simulation
program.  Thus, this measurement was also an important check of the
correctness of the simulation program.

\section{Simulation program and parameters}
\label{sc:simulations}

The equations of motion (\ref{eom}), (\ref{gauge-eom}) are discretized
as described in Ref.~\cite{lmax}; we invite interested readers to
check therein for the detailed implementation.  We note that we
implement the $W$-fields in a ``staggered'' fashion: because the $W$
equations of motion only have first order derivatives, a symmetric
discretization decouples $W$-fields at even and odd lattice sites from
each other (i.e. at space-time sites where the integer coordinate
$n_x+n_y+n_z+n_t$, where $n_t$ is the number of the evolution time
step, is either even or odd.).  Thus, we can delete the $W$-field at
odd sites, saving memory and cpu-time.%
\footnote{This procedure also deletes half of the unphysical doublers
  inherent in the $W$-field spectrum.  The reason these doublers
  appear is the same as for the notorious lattice fermion doublers,
  namely the first order derivatives.  However, in our case the
  doublers are quite benign, as is discussed in \cite{lmax}.}

For the time update we use a time-symmetric staggered leapfrog as
described in \cite{lmax}.  The only essential difference is the 
appearance of the last term in Eq.~(\ref{eom}).  In order
to guarantee that the update remains invariant under time reversal
we implement the update of the $W$-fields in two stages, interleaving these 
with the gauge and electric field update steps.

The time-step values we use are $\delta t = 0.05 a$ and $0.1 a$, where
$a$ is the spatial lattice spacing.  We shall discuss the lattice
artifacts -- finite $a$, finite volume, finite $\delta t$, and finite
$\lmax$ -- in detail in Sec.~\ref{sc:lattice}; to summarize, all 
lattice effects appear to be well under control.

We note that while all $m_l^2$ are dimensionful in the equations of
motion, for fixed asymmetry the ratios $m_l^2/m_0^2$ remain constant.
Thus, every dimensionful quantity can be given in terms of the powers
of single parameter, $m_0^2$.  In particular the lattice spacing is
given as $(a m_0)$.  The gauge coupling constant $g^2$ can be
completely absorbed in the equations of motion, making the results
independent of the value of $g^2$.

Our initial conditions are as follows: we initialize the electric
field components $\vec E^a(\vec x)$ to a small amplitude white noise,
i.e. random Gaussian fluctuations, with vanishing initial $\vec A $
and $W_{l m} $.  We make an orthogonal projection of the $E$-fields to
a hypersurface satisfying Gauss' law, $D_i E^i = 0$ (since
$W_{00}=0$).  The evolution equations preserve Gauss' law.  The
electric field drives the gauge field $\vec A$ to a non-zero value
very quickly, so that $\langle \vec B^2 \rangle \approx \langle \vec
E^2 \rangle$ before the exponential growth of the unstable modes
becomes visible.  The amplitude of the initial fluctuations is chosen
small enough so that the equations of motion are essentially linear
during the initial stage.  The growth of unstable modes then drives
the fields to much larger values.

The lattice spacings and sizes used in the analysis 
are shown in table~\ref{tab:lattices}.  The simu\-lations have been
performed mostly using pc-clusters with infiniband interconnects.  
The simulations require unusually large amounts of memory 
(for lattice simulations); our largest simulations used 192 nodes, 
with a total memory requirement of around 400\,GB.  The simulations
were performed at the Finnish IT Center for Science (CSC).

\begin{table}
\begin{center}
\begin{tabular}{l|l|l}
\hline
$m_0 a$  & $\lasym=2$, $\lmax=16$ & $\lasym=4$, $\lmax=16$   \\
\hline
  1      & $64^3$     & $64^3$           \\
  0.77   & $64^3$     & $64^3$           \\
  0.55   & $64^3$, $80^3$, $104^3$, $120^3$ 
                      & $64^3$, $128^3$  \\
  0.45   & $64^3$, $120^3$ & $160^3$     \\
\hline
\hline
$m_0 a$  &$\lasym= 14$, $\lmax=16$  & $\lasym= 28 $, $\lmax=32$ \\
\hline
  1      & $64^3$ &  $48^3$    \\
  0.77   &        &  $64^3$    \\
  0.71   & $64^3_{\bf 16,24}$ &            \\
  0.55   & $48^3$, $64^3_{\bf 16,24}$, $96^3$, $128^3$  & $64^3$ \\
  0.32   & $64^3_{\bf 16,24}$, $80^3$, $96^3$, $120^3_{\bf 16,24,32}$, 
  $180^3$
            &      \\
  0.30   & & $64^3$, $96^3$, $128^3_{\bf 32,48}$, $192^3$ \\
  0.17   & $240^3$ & $180^3$   \\
  0.10   & $240^3$ & $240^3$   \\
\hline
\end{tabular}
\end{center}
\caption[a]{
The lattice spacings (in units of $m_0$) and lattice sizes used in the 
weak initial field analysis for each value of the asymmetry.  For
several of the volumes there are more than one individual run.
The $\lmax$-cutoff used is shown at the top of the columns.
In addition, there are some some volumes with more than one 
$\lmax$-cutoff; these are indicated with a subscript
(only for $\lasym=14,28$).
}
\label{tab:lattices}
\end{table} 


\section{Results}
\label{sc:results}

\subsection{Energy densities}

\begin{figure}[t]

  \vspace*{0.5cm}

  \centerline{
              \epsfxsize=8cm
              \epsffile{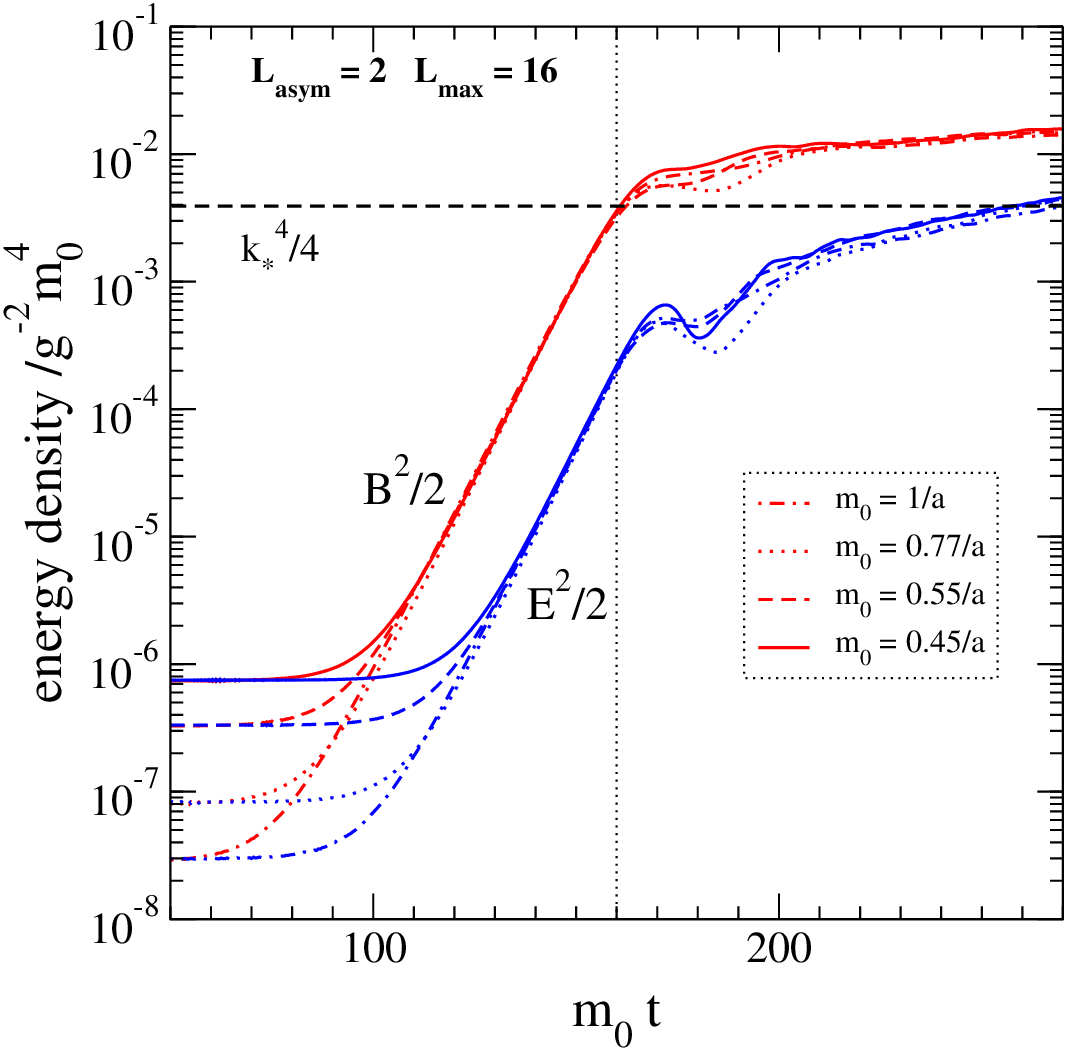}
              \epsfxsize=8cm
              \epsffile{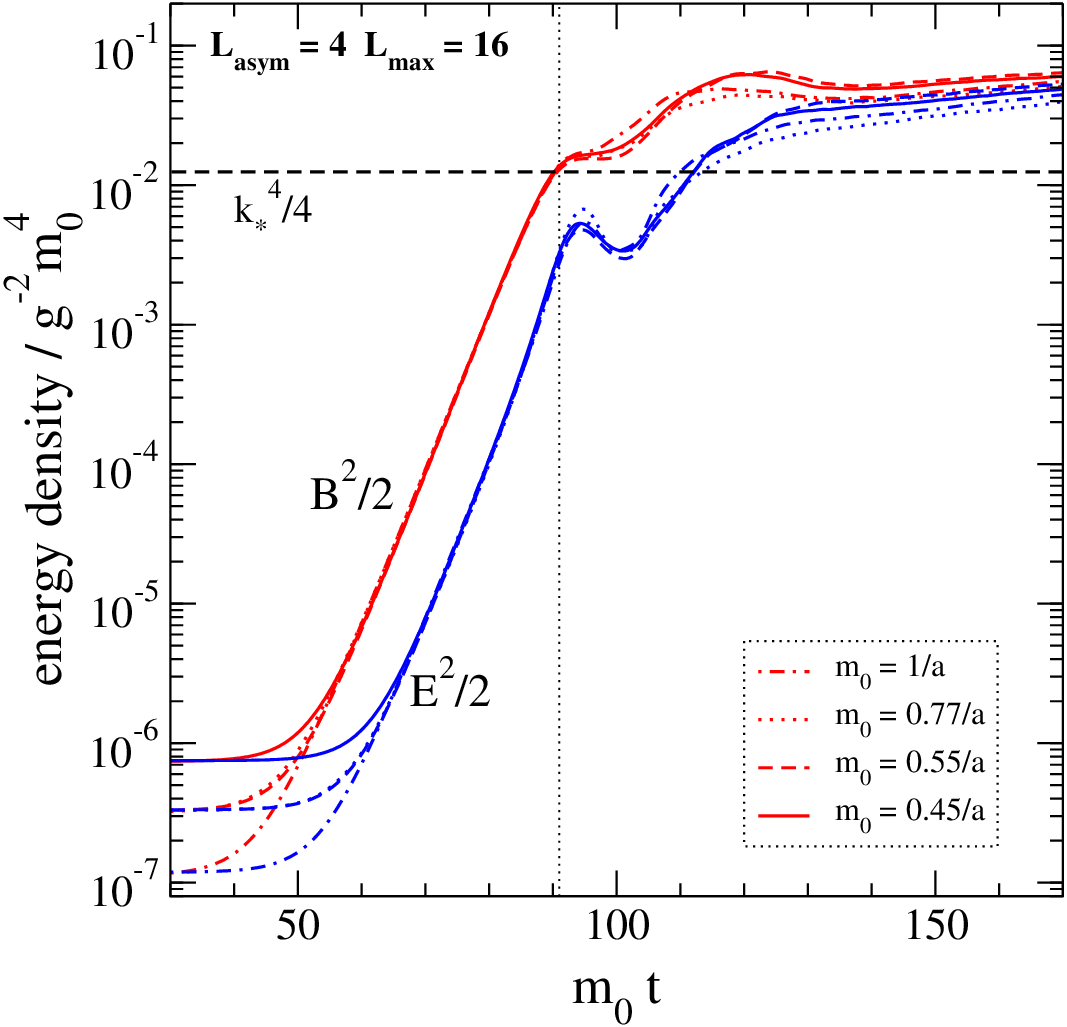}
             }

             \caption[a]{Magnetic and electric field energy densities
               as a function of time for moderate anisotropy, measured
               from lattices with different lattice spacings $a$.  The
               lattice sizes are the largest ones for each lattice
               spacing in table~\ref{tab:lattices}.  }
  \label{fg:moderate}
\end{figure}

As mentioned in Sec.~\ref{sc:simulations}, the initial condition we
use is a white noise spectrum satisfying Gauss' law for the electric
field, with $\vec A$ and $W$ set to zero.  In Fig.~\ref{fg:moderate} we
show the field evolution for weak anisotropy ($\lmax=2,4$) starting
from very small amplitude initial conditions for different values of
the lattice spacing.  We see qualitatively the same behavior as
observed in Refs.~\cite{arnoldFate,romatschkeFate}.  After some
initial settling down, the soft fields start to grow exponentially
until they reach the non-abelian point $ \vec A \sim \vec k /g $ where
non-linear terms in the equation of motion start playing a role.  We
find that this happens when the magnetic field squared approximately
equals
\begin{eqnarray}
    \frac12 \vec B ^2  _{\rm non-abelian} 
    \simeq  \frac{ k _ \ast^4 }{4 g^2}
  .
     \label{nonabelian}
\end{eqnarray} 
After that  the growth
slows down significantly and is no longer exponential. According to Arnold and
Moore \cite{arnoldCascade} this growth is due to cascading of energy from the
originally unstable infrared modes to higher momentum ones. 
The amplitude of the initial fields was not specifically tuned to be equal 
for different lattice spacings; nevertheless, the gauge field evolution
falls on a curve independent of the lattice spacing (as long as the
volume is large enough, see Sec.~\ref{sc:lattice}).  The origin of time 
$t=0$ has been adjusted in Fig.~\ref{fg:moderate} in order for the 
growth phases to overlap.  Thus, only differences of $t$ have a physical
meaning.

\begin{figure}

  \vspace*{0.5cm}

  \centerline{
              \epsfxsize=8cm
              \epsffile{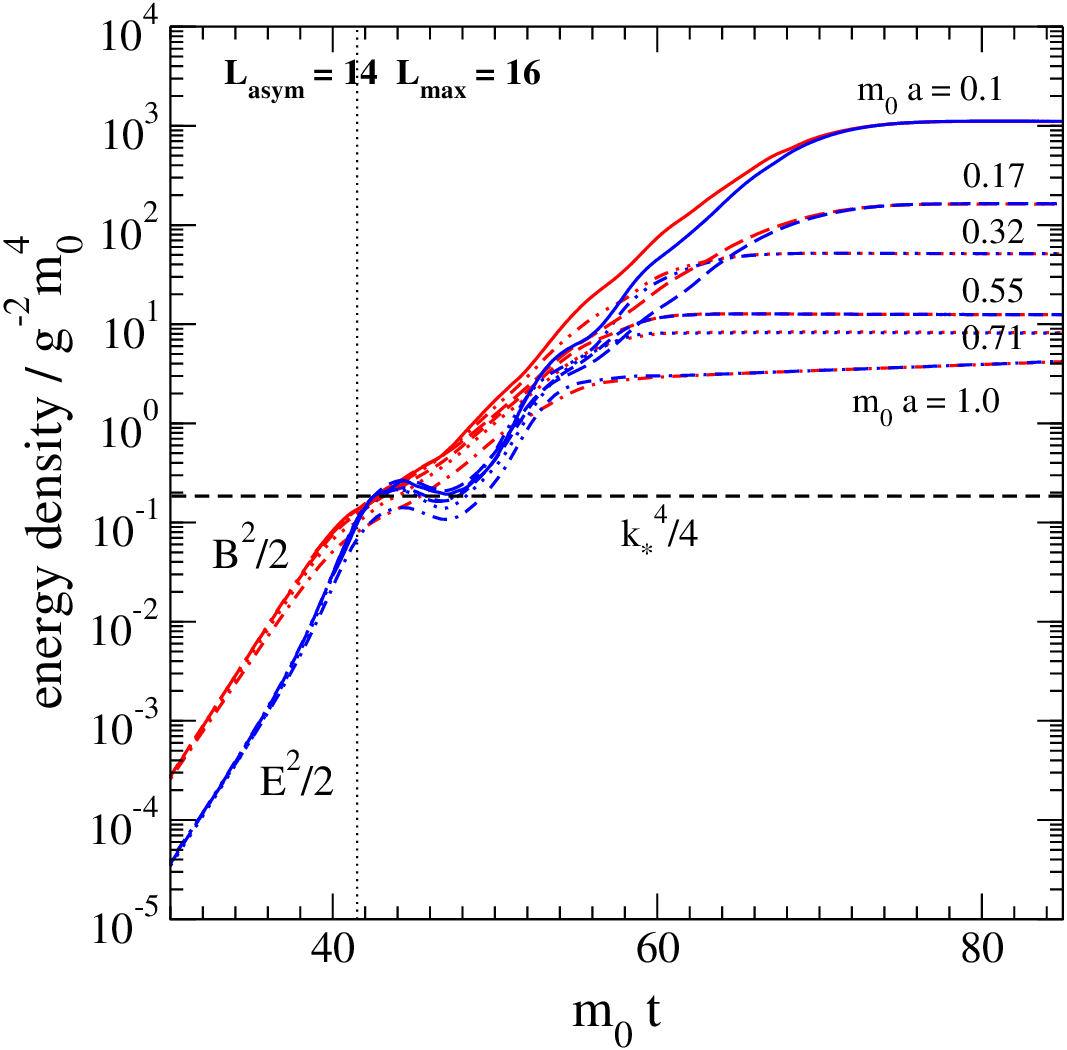}
              \epsfxsize=8cm
              \epsffile{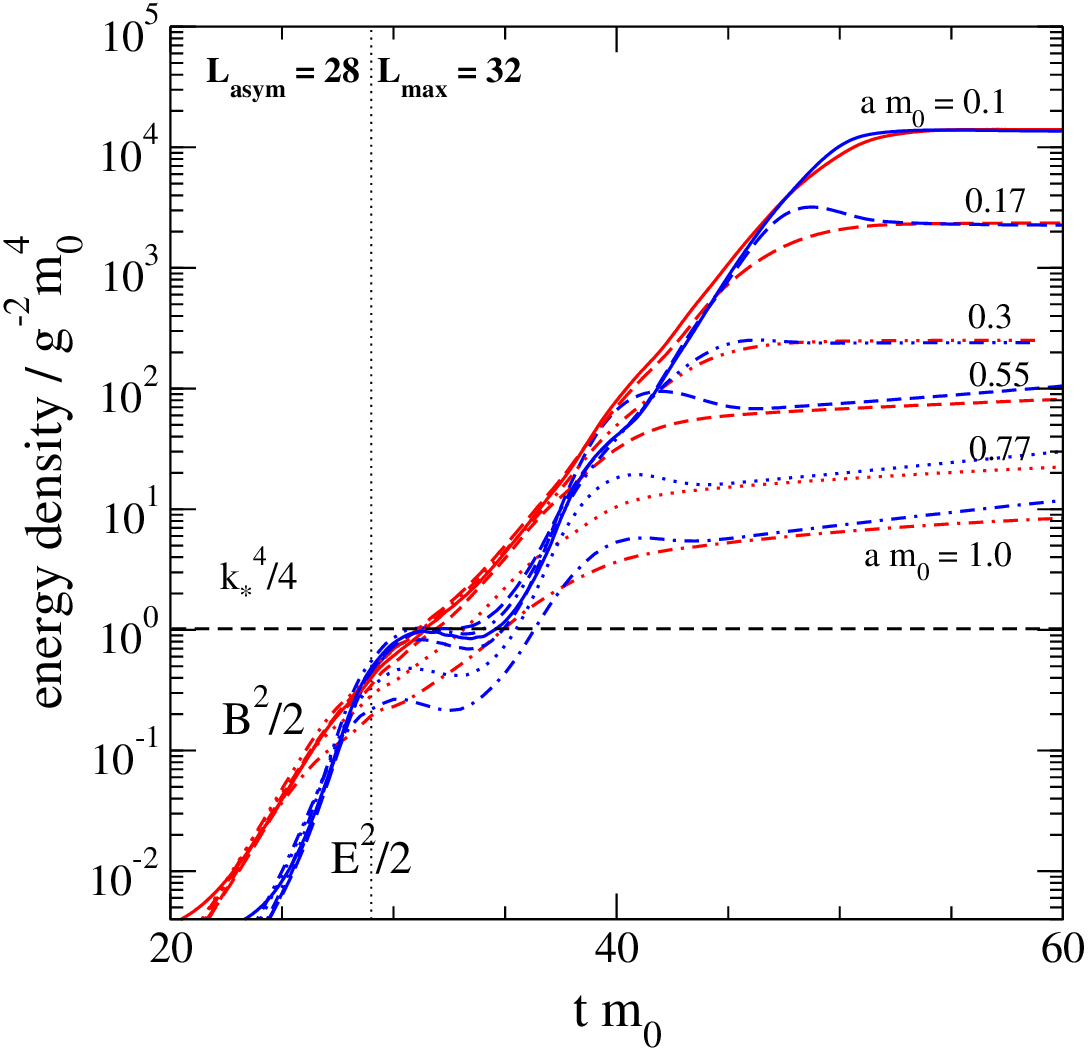} 
             }
             \caption{Same as Fig.~\ref{fg:moderate} but for stronger
               anisotropy. Now the growth of field energy appears to
               continue indefinitely and it is stopped only by lattice
               cutoff effects.  For each lattice spacing we show the
               largest volume listed in table~\ref{tab:lattices}.
             }
  \label{fg:strong}
\end{figure}

For strong anisotropies we find a very different picture. In
Fig.~\ref{fg:strong} we show our results for $\lasym = 14$ and $28$.
We clearly see the onset of non-linear effects at the magnetic field
energy density around $k_\ast ^ 4 /(4 g ^ 2) $.  There the growth
ceases to be exponential and the dynamics becomes very complicated.
The electric field grows very rapidly, and the electric field energy
becomes as large as the magnetic one. Subsequently, however, the
growth of energy continues at a large rate. It is not a purely
exponential growth, but it is not much slower than the initial weak
field growth. For $ \lasym = 28 $ the growth rate is roughly as large
as in the weak field regime ($ m _ 0 t < 40 $).  

At some value of the energy density the growth saturates.
Furthermore, in contrast to the moderate asymmetry in
Fig.~\ref{fg:moderate}, the electric and magnetic field energies
reach an equal level at the end.  
In Fig.~\ref{fg:strong} we show the values where the growth
finally saturates for different values of the lattice spacing $ a $.
We see that the
saturation energy has a strong dependence on the lattice spacing, growing
as $a m_0$ is decreased.
Therefore we can conclude that the 
saturation seen in Fig.~\ref{fg:strong} is caused by the
lattice regularization.

\begin{figure}[t]

  \vspace*{0.5cm}

  \centerline{
  \epsfxsize=8cm
  \epsffile{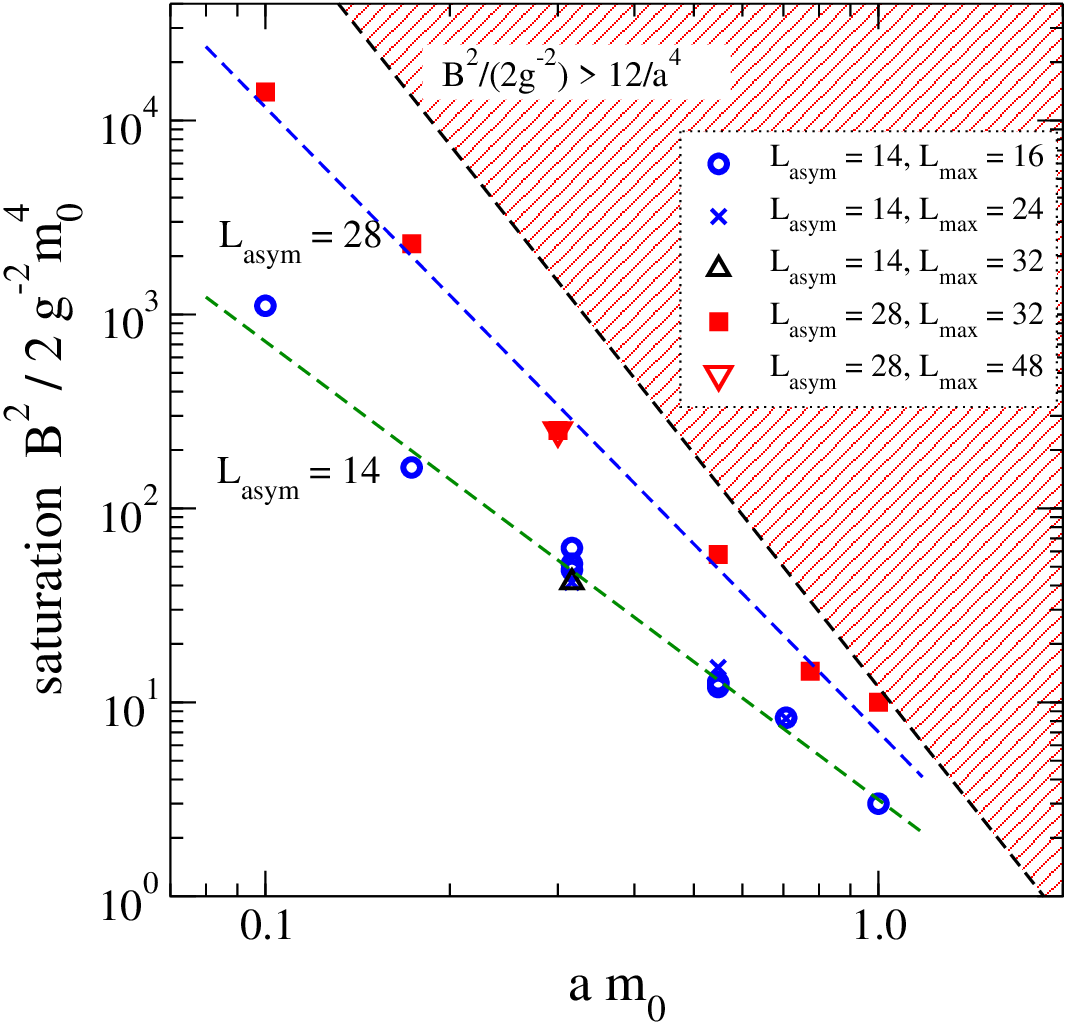}
    }

  \caption{Maximal magnetic field energy density  as a function of the
    lattice
    spacing for $\lasym=14$ and $28$, and for all values of $\lmax$
    used.
    The shaded region
    is above the maximum magnetic field energy density, given by a
    completely random lattice gauge system.  The dashed lines are
    power-law fits to the two asymmetries, with the results
    $(am_0)^{-2.4}$ ($\lasym=14$) and $(am_0)^{-3.2}$ ($\lasym=28$).
  }
  \label{w3d_L14L28_saturation}
\end{figure}

In Fig.~\ref{w3d_L14L28_saturation} we show the maximal magnetic
energy density as a function of the lattice spacing. The maximal
energy density appears to grow without
bound with decreasing $ a $ with a power-like behavior. 
The magnetic field energy on the lattice is given by 
$ 4/( a g ^ 2 ) \sum _ { i < j } 
  ( 1 - \frac 12 {\rm Tr}\, U _ { i j } ) $, where 
$  U _ { ij }  $ 
is the ordered product of link variables around a spatial plaquette,
\begin{eqnarray} 
    U _ { ij }    ( x ) 
  \equiv U _ i ( x ) U _ j  ( x + a  {\bf \hat  i }  ) U ^\dagger _ i ( x + a
   {\bf \hat  j } ) U ^\dagger _ j ( x ) .
\end{eqnarray}  
There is an absolute upper limit on the magnetic energy density,
$24/(a^4g^2)$ which is  reached when ${\rm Tr}\,U_{ij} = -2$.  This is a very
particular fully ordered state; a more realistic limit is the
completely random state where $\langle {\rm Tr}\, U_{ij} \rangle = 0$
and where the magnetic energy density reaches the limit $12/(a^4g^2)$.  Energies
above this limit are shown in Fig.~\ref{w3d_L14L28_saturation} 
as a shaded region.  

We observe that our maximal field energies do not quite reach the
maximum energy limit.  Instead, the saturation energy density appears
to diverge in the continuum limit with a different power of $a$.  If
we fit a power law behavior to the saturation energy density at both
asymmetries, we obtain the results $E_{\rm saturation} \propto
(am_0)^{-2.4}$ for $\lmax=14$ and $(am_0)^{-3.2}$ for $\lmax=28$.
Because we do not have proper statistical errors for the data in
Fig.~\ref{w3d_L14L28_saturation}, we cannot quote proper error bars
for the fitted exponents.  However, we can nevertheless make a rough
estimate of them by performing jackknife analysis in terms of the
individual simulation points, obtaining an error bar $\pm 0.2$ for both
exponents.  It is worth noting that the exponent in the $\lasym=28$
case is close to $-3$, the exponent given by the thermal distribution
with a lattice cutoff.

This analysis shows that there appears to be no
saturation of the energy density if the lattice spacing is removed.  
This is very different
behavior from the one that was observed in the 3+1 dimensional
simulations of Refs.~\cite{arnoldFate,romatschkeFate}.

Let us now discuss possible reasons for this behavior. 
When the
anisotropy is mild, the unstable modes have momenta of order $ m _ 0 
$.  However, for strong anisotropy there are unstable
modes with $  |  k _\perp  | \lsim
m _ 0 $ but with longitudinal momentum $  |  k _  z  | $ all the way up to $
k _ { \rm max } $,   where  
\begin{eqnarray} 
       k _{\rm max}
       \sim  \frac{ m _ 0 }{\eta  } 
    \label{kmax} 
\end{eqnarray}
and $ \eta $ is the measure of anisotropy introduced in eq.~(\ref{eta}).
In \cite{arnoldTurbulent} it was argued that the magnetic field squared of
these modes cannot become larger than $ \vec
B ^ 2 \sim m _ 0 ^ 4/(g ^ 2 \eta  ^ 2)$. 
The energy density at saturation in a strongly anisotropic
plasma would then be enhanced by a factor $ 1/ \eta    ^ 2 $ compared to the
case of moderate anisotropy. However, this enhancement factor is only about $
16 $ for $ \lasym = 14 $ and about 67 for $ \lasym = 28 $, 
while we see the energy density
in Fig.~\ref{fg:strong} growing by many orders of magnitude larger than
in the case of weak anisotropy. Therefore it is not a (quasi-) exponential
growth of modes who's equations of motion are almost linear which 
could explain the behavior seen in Fig.~\ref{fg:strong}. 
Thus the continued growth must be an effect which is essentially
non-linear. 

There appear to be (at least) two scenarios for the physics
behind the continued growth. The first is
that the unstable modes  grow to occupancy much
larger than $ 1/g ^ 2 $ as suggested in
Ref.~\cite{arnoldAbelianization}. 
Another possibility is that the energy goes into the high momentum
modes, rather than into the modes which are unstable in the weak field
regime.  We shall try to distinguish between these outcomes by measuring
quantities which are sensitive to the momentum spectrum of the gauge fields:
gauge fixing and direct Fourier transformation, gauge invariant operators
and gauge invariant cooling.  These all indicate that the energy
indeed gets dumped to the UV, and there is no growth of the IR modes
much beyond the non-abelian point.

\subsection{Coulomb gauge occupation numbers}
\label{sc:coulomb}

For free gluon fields the concept of  occupation
numbers $ f _ {\rm s}  (\vec k ) $\footnote{We use the subscript s to
  distinguish the occupation number of the classical (soft) fields from the
  occupation number of hard gluons which are described by the $ W $-field.}
is unambiguous. It can be calculated from the gauge field by fixing to
Coulomb gauge using the expression 
\begin{eqnarray}
  f _ {\rm s}   ( \vec k ) = \frac{  |  \vec k  |  } {2 V N _{\rm dof} } \left  |  \vec
  A ( \vec k )  
  - \frac{ i } {  |  \vec k  |  } \vec E ( \vec k ) \right |  ^   2 
  \label{occupancy}
\end{eqnarray}
where $ N _{\rm dof} $ denotes the number of color/spin degrees of freedom. 
For reflection invariant field configurations the interference term of $ \vec A $ 
and $ \vec E  $ vanishes. For free fields the two remaining terms
give equal results when they are averaged over time.  
Thus, assuming reflection
invariance, one can compute the occupancy either from $ \vec A $ or from 
$ \vec E $ only, and in this work we use the former case. 
The distributions shown here are averaged over all directions of $\vec k$,
\begin{eqnarray} 
   f (k) \equiv   \int \frac{ d\Omega } { 4\pi  } f _ {\rm s} (\vec k)
\end{eqnarray} 

If the gluon field amplitudes are large and/or the gluons are
interacting with the particles, there is no occupation number in the
strict sense. Nevertheless one expects that (\ref{occupancy}) still
gives a good estimate of the power in one field mode.  However, fixing
the gauge for large fields in a non-abelian theory is dangerous due to
Gribov copies of near vacuum configurations of the high momentum
modes.  We make three consistency checks of the gauge fixed spectrum
by comparing with gauge invariant measurements: the total energy in the
gauge fixed spectrum, measurement of the average $\langle \vec k^2\rangle$,
and comparing the spectrum with gauge invariant cooling.  These will
be discussed below.

\begin{figure}[t]
  \vspace*{0.5cm}

  \centerline{
              \epsfxsize=8cm
              \epsffile{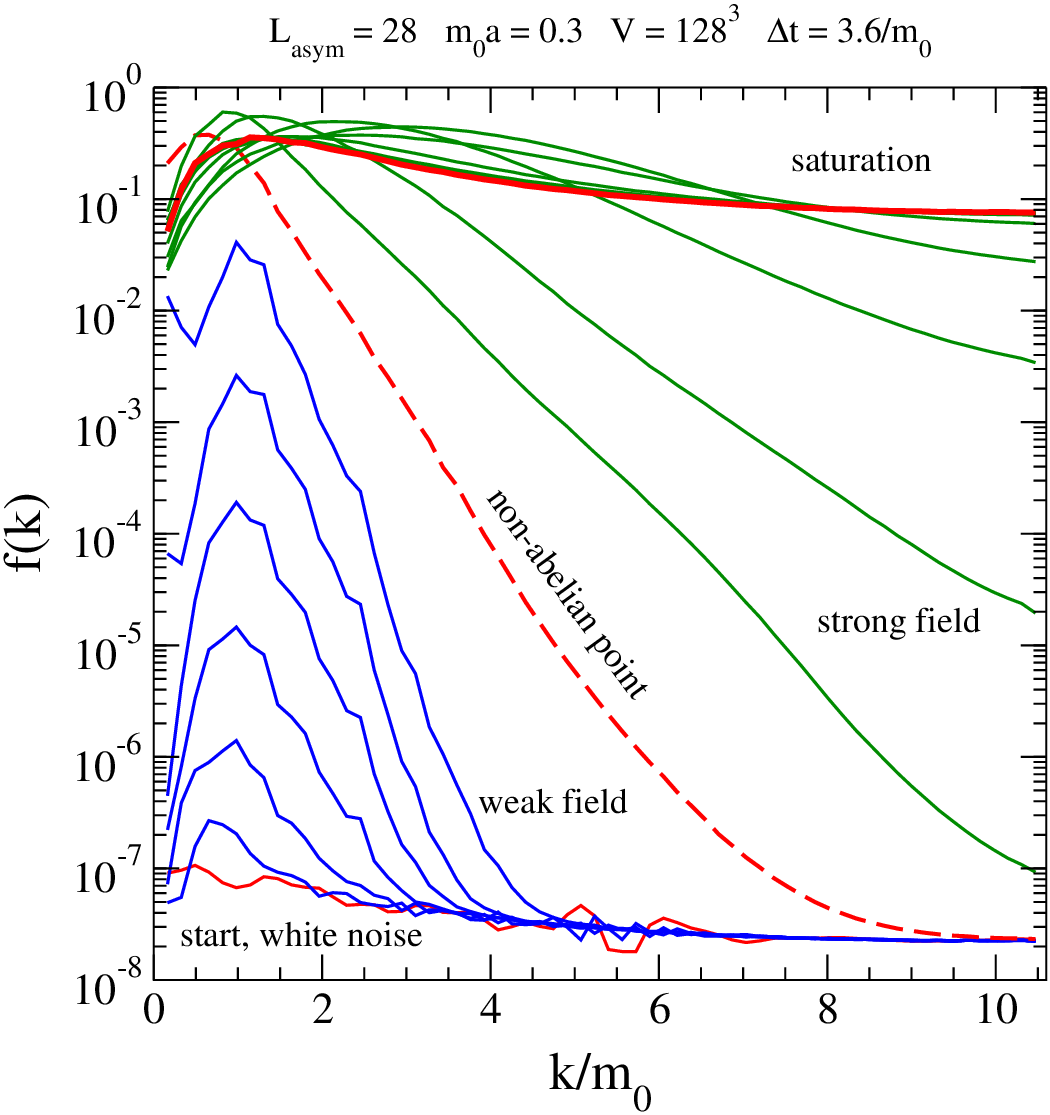}
              \epsfxsize=8cm
              \epsffile{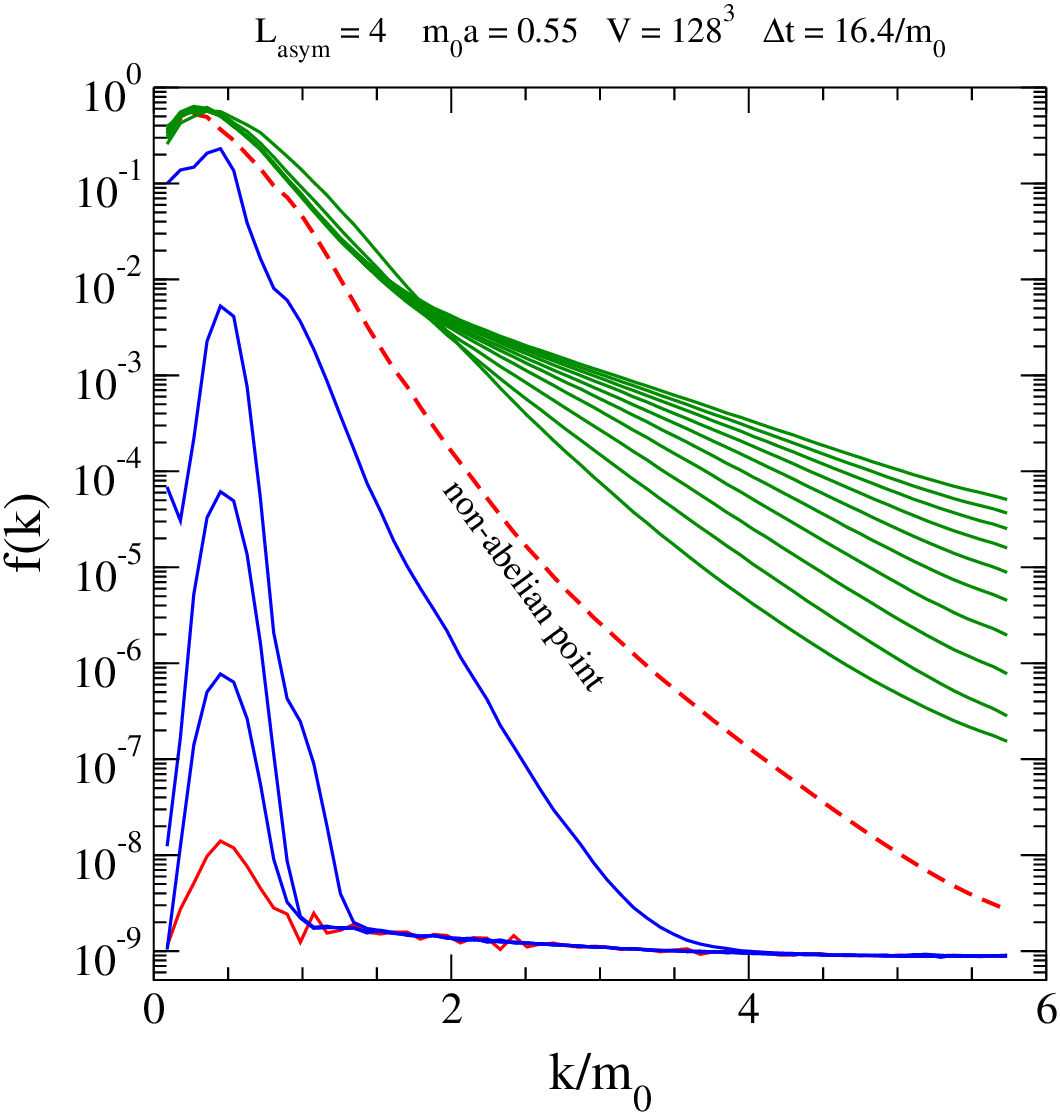}
             }

             \caption{Coulomb gauge power spectrum (occupation number)
               as a function of time for strong ($\lasym=28$, left)
               and weak ($\lasym=4$, right) anisotropy.  The power
               spectra are plotted at equal intervals of $\Delta t = 3.6/m_0$
               for $\lasym=28$ and $\Delta t = 16.4/m_0$ for
               $\lasym=4$.  }
  \label{fg:occupation}
\end{figure}

The occupation numbers as a function of time are shown in
Fig.~\ref{fg:occupation}, for strong ($\lasym=28$)
and moderate anisotropy ($\lasym=4$).  
The curves show the spectrum measured at constant evolution time intervals.  
Early times are at the bottom; 
the initial white noise $\vec E$-field implies a spectrum 
$f(k) \sim 1/k$.

Let us first consider the case of strong anisotropy.  
At early times one sees a rapid growth of
the infrared modes which is the fastest at $ k = k _ \ast $. The
dashed curve is at the time at which non-linear effects become
important.  In Fig.~\ref{fg:strong} this time 
is marked with
a vertical dotted line.  As this is happening the active mode
spectrum widens very rapidly.  At later time times the amplitude of the
$k \sim k_\ast$-modes does not grow any longer, but the ultraviolet end of the
spectrum grows extremely rapidly -- in fact the occupation
number at higher $k$ grows faster than the original growth rate 
at $k_\ast$, as can be observed from the large gaps between the lines in 
Fig.~\ref{fg:occupation}.  The final spectrum is shown with a thick
line, and its shape fits $f(k)  \sim 1/k$ quite well,
consistent with a thermal distribution.  However, a more detailed inspection
of the spectrum shows that the growth of the energy stops before this
is reached: the growth stops when the occupation numbers near the
lattice cutoff $k/m_0 = \pi/(m_0 a) \approx 10.5$ become
appreciable ($\gsim 0.05$).  After this the distribution just
settles towards the thermal one, without increase in energy.

The situation at modest anisotropy (Fig.~\ref{fg:occupation} right)
looks quite similar at the beginning.  However, in this case the
growth in the UV part of the spectrum stops soon after the non-abelian
point is reached.  The mode spectrum remains dominated by the IR
modes, and the total energy grows only approximately linearly with
time.

\begin{figure}[t]

  \vspace*{0.5cm}

  \centerline{
              \epsfxsize=8cm
              \epsffile{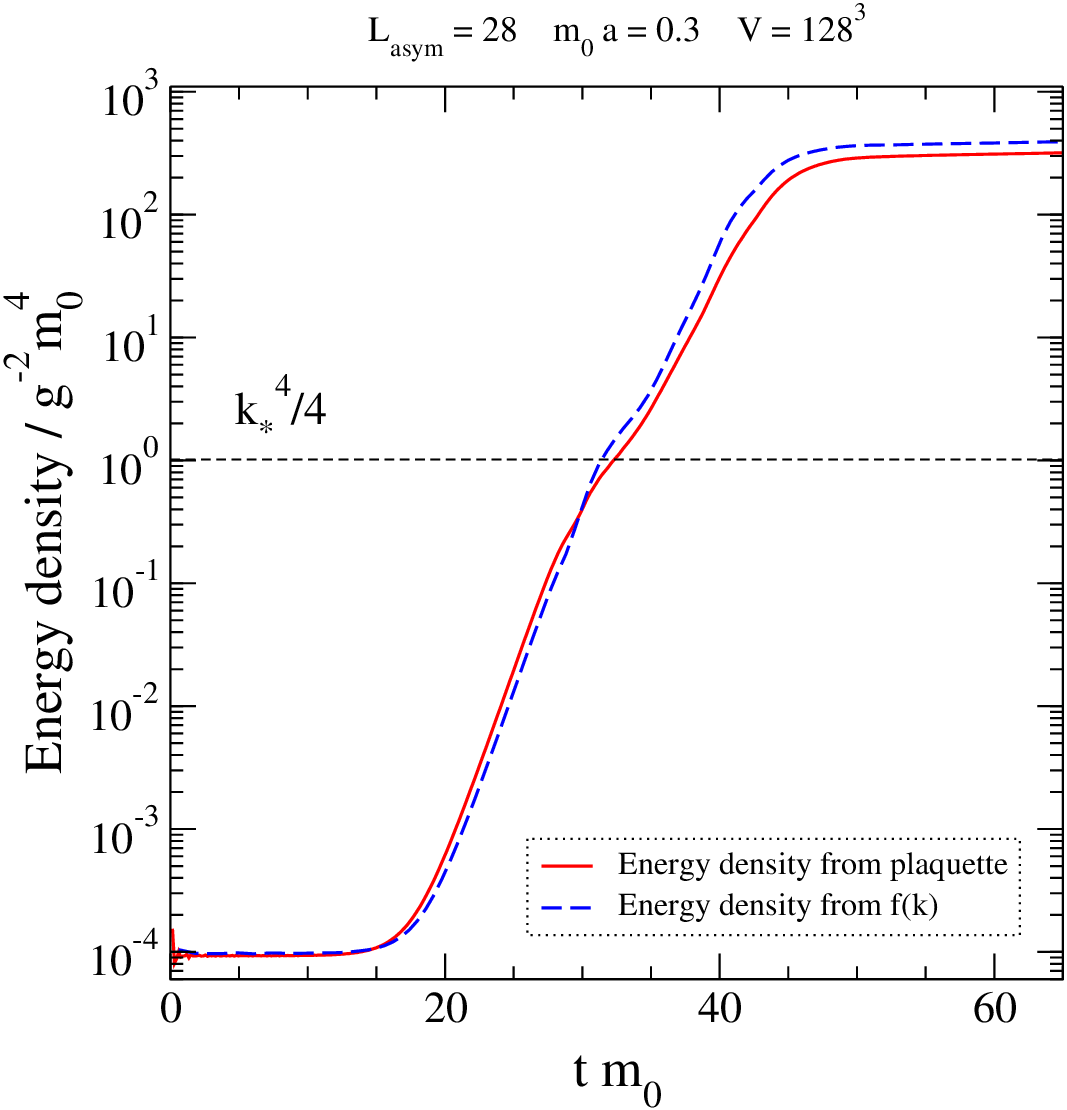}
             }
             \caption{Energy density computed from the Coulomb gauge 
               power spectrum,
               compared with the true energy density in the magnetic field,
               for $\lasym=28$, $m_0a = 0.3$.
             }
             \label{fg:energy}
\end{figure}

As a check that the occupation number reflects the true distribution of
energy over the different modes we compute the total field energy
density $ \varepsilon $
\begin{eqnarray}
  \varepsilon  = \int \frac{ d ^ 3 k } { ( 2 \pi  ) ^ 3 }  |  \vec k  | 
  f _ {\rm s}   ( \vec k ) = \frac1{(2\pi)^3}\int_0^\infty dk\,k^3 f(k)\,,
  \label{free}
\end{eqnarray} 
and compare it to the
(gauge invariant) direct measurement of the energy from the lattice.  
The result is shown in
Fig.~\ref{fg:energy}.  In the weak field regime our measured $ f $
slightly over-estimates the energy density. One has to keep in mind
that even for very small amplitudes the gauge fields are not free, but
are coupled to the $ W $-fields, so that the two curves need not
coincide exactly. At large fields the discrepancy is slightly bigger
and $ f $ yields a slightly too large result.
However, the overall 
disagreement is within a factor of 1.4.

\subsection{Average $  |  \vec k  |  $ from gauge invariant operators}

\begin{figure}[t]

  \vspace*{0.5cm}

  \centerline{
              \epsfxsize=8cm
              \epsffile{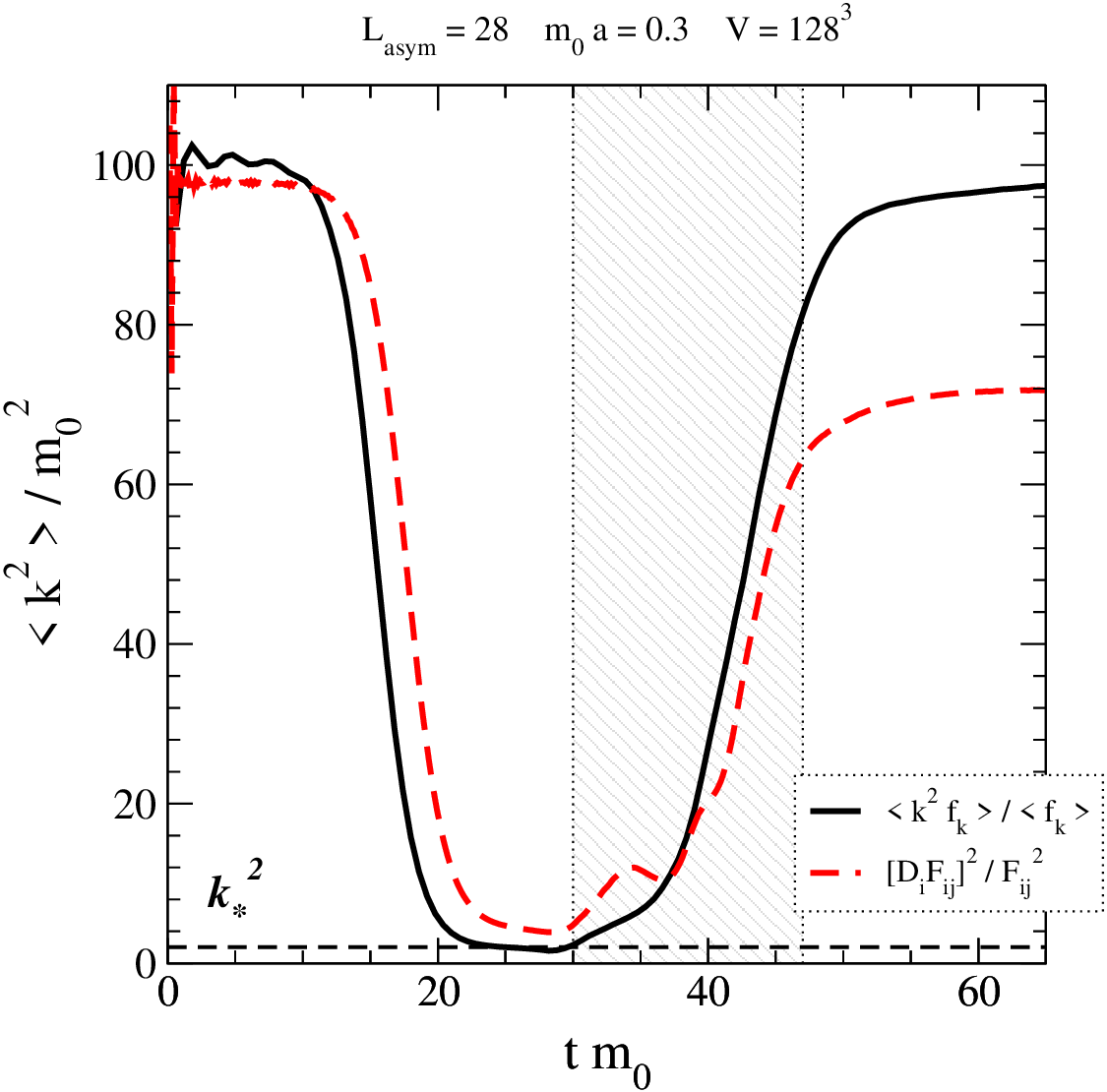}
             }

  \caption{Average $ \vec k ^ 2 $ as a function of time, measured from
    the gauge fixed occupation numbers $f(k)$, and from the gauge
    invariant operator, Eq.~(\ref{averagek2}), for the $\lasym=28$,
    $m_0 a=0.3$ -simulation shown in Figs.~\ref{fg:occupation} and
    \ref{fg:energy}.  The shaded region is the time interval when
    the non-linear growth of energy is occurring.
  }
  \label{fg:averagek2}
\end{figure}

The Coulomb gauge occupation numbers strongly indicate that the continued
growth seen above is due to population of high momentum
modes. However, one may  be concerned about gauge artifacts, because strong
fields could produce fake high momentum occupancy. Therefore, in order to be
certain about our conclusion regarding the high momentum occupation, 
it is mandatory to investigate 
this result also using gauge invariant measurements. 
A measure for the typical
momentum squared of the color-magnetic  fields is 
\begin{eqnarray} 
   \langle \vec k ^ 2 \rangle \equiv  \frac{ \int {\rm tr} 
   ( \vec D      \times \vec B ) ^ 2 d ^ 3 x } { \int {\rm tr} \vec B ^ 2 d ^
   3 x } 
   \label{averagek2}
\end{eqnarray} 
In electrodynamics this would equal
\begin{eqnarray} 
    \langle \vec k ^ 2 \rangle _ { \rm QED } =  \frac{ \int  
    \vec k ^ 2 |  \vec B ( \vec k )   |   ^ 2   d ^ 3 k } 
  { \int   |  \vec B  ( \vec k )  |  ^ 2 d ^    3 k} 
  \nonumber 
\end{eqnarray} 
In QCD there is also the commutator $ [  A _ i , B _ j ] $ contributing to $
\langle \vec k ^ 2 \rangle  $. So it   appears that large  $ \langle \vec k ^ 2
\rangle $ does not necessarily imply that the typical $ \vec k ^ 2 $ of the magnetic field
is large. However, in the 1-dimensional simulations 
\cite{romatschke2d} where the unstable modes grow indefinitely, the commutator
terms were found to remain small in accordance with the abelianization picture
of Ref.~\cite{arnoldAbelianization}.
Thus we expect our $ \langle \vec k ^ 2
\rangle $ to be a good measure of the momentum of the modes. Note in
particular that the commutator term is parametrically of the same size as the
gradient term when non-linear effects start playing a role. 

In Fig.~\ref{fg:averagek2} we show $ \langle \vec k ^ 2 \rangle $ as a function
of time, both computed from the gauge invariant object (\ref{averagek2}) and
from the Coulomb gauge 
occupation numbers. At early times ($ t < 12/m _ 0 $),  
when the fields are very weak, $ \langle \vec k ^ 2 \rangle $ is large because
it is dominated by UV
modes due to our white noise initial conditions. As soon as the unstable modes
start growing they give the dominant  contribution to $ \langle \vec k ^ 2
\rangle $  which is then of order $ k _\ast ^ 2$.  
The two curves do not coincide which is not surprising
since even for free fields they would in general not be identical. 
Once one is in the
non-linear regime, the average $ \vec k ^ 2 $ increases rapidly. This is a
clear signal of a rapid transfer of energy to high momentum field modes. When 
the lattice cutoff starts having an influence on the time evolution ($ t m _ 0
\gsim  45 $), the two curves start to deviate strongly. 

\subsection{Cooling}

\begin{figure}[t]

  \vspace*{0.5cm}

  \centerline{
    \epsfxsize=10cm
    \epsffile{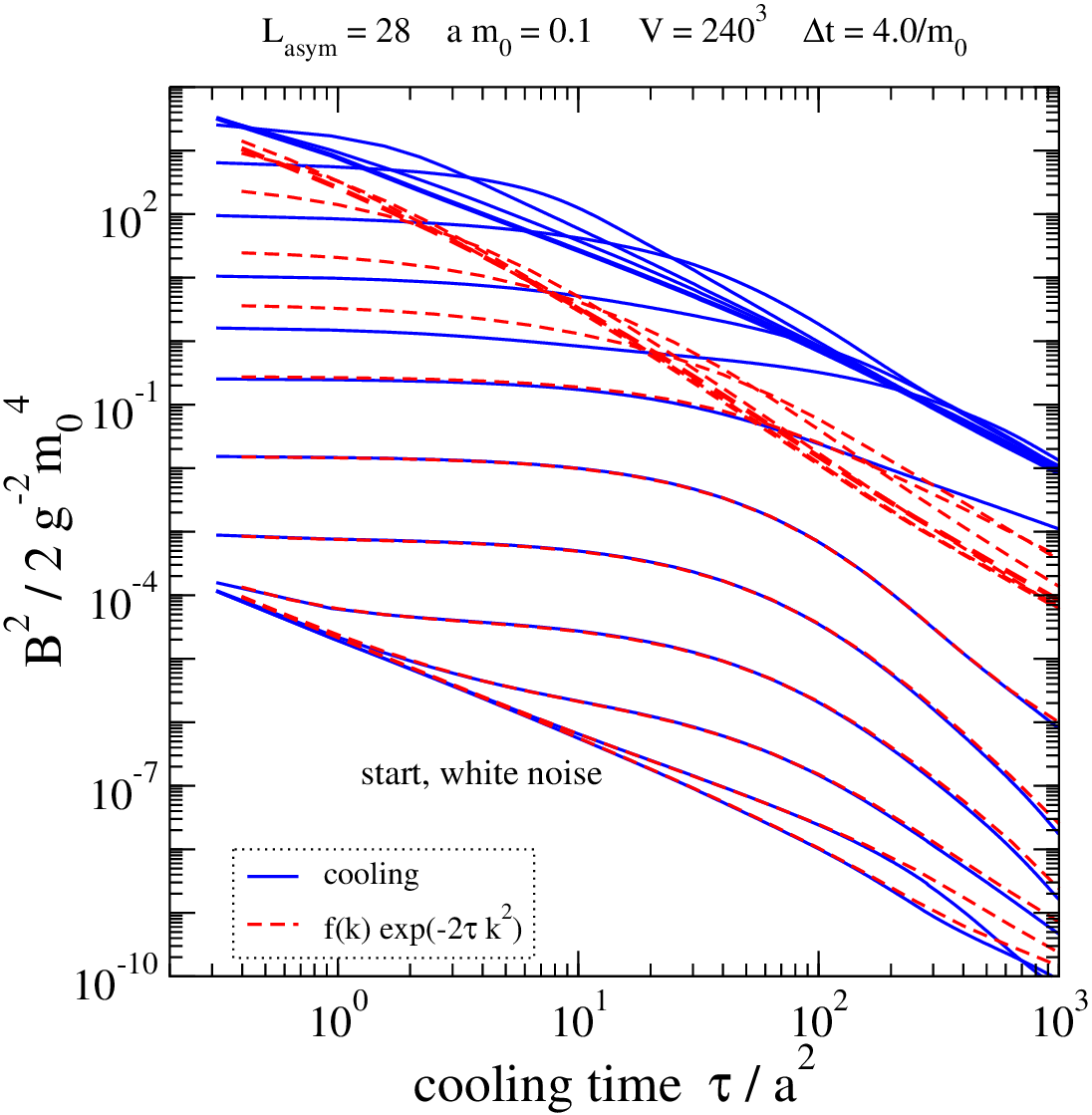}
  }
            
  \caption{Magnetic field energy (solid lines) during the cooling of
    the field configurations. The dashed lines are obtained from the
    `cooled' Coulomb gauge occupation numbers, Eq.~(\ref{fcool}). The
    different curves are for physical times in intervals of $ 4/m_0 $,
    with time increasing from bottom to top.  The final curves for
    both cases are shown with thicker lines.  }
  \label{fg:cooling}
\end{figure}

Another gauge invariant method for obtaining  
information about the gauge field spectrum 
at a given physical time
is to take the gauge field configuration at that  
 time and let it evolve in the (unphysical) cooling 'time' $ \tau  $ using the
equation of motion 
\begin{eqnarray}
\partial_\tau A_i = D_j F_ { ji }
\end{eqnarray} 
This reduces the gauge field energy monotonously.
For weak fields the Fourier components in Coulomb gauge evolve like 
\begin{eqnarray}
   A _ i ( \tau  , \vec k ) = \exp( - \tau  \vec k ^ 2 ) A _ i( 0, \vec k ) .
   \label{cool}
\end{eqnarray}
Thus, the cooling has the largest effect on the high momentum modes
and they are depleted first.
 Results for the cooling time dependence
of the magnetic field energy are shown in Fig.~\ref{fg:cooling} (full
lines), measured at intervals $\Delta t = 3.6/m_0$ during the evolution of a
system with strong anisotropy ($\lasym=28$).

For free fields with a thermal spectrum Eq.~(\ref{cool}) gives the
result $\mbox{Energy} \sim \tau^{-3/2}$ for large enough $\tau$.  
This behavior is clearly
visible at early time cooling curves, the bottom curves in
Fig.~\ref{fg:cooling}.%
\footnote{Our initial condition was small amplitude white noise for
  $\vec E$, which is thermal by itself.  This rapidly populates $\vec A$-modes
  to an approximately thermal distribution.}
When we are in the linear regime where the unstable modes grow exponentially, practically
all of the 
the energy is in the infrared modes, and the cooling takes 
more time to have any effect on the total energy.  
This is visible as horizontal lines
in the middle part of the cooling plot.  
When the cooling time reaches $\tau \sim 1/k_\ast^2$, the
energy starts to decrease rapidly and the cooling curves
develop a smooth shoulder.

The results from the gauge invariant cooling can be directly 
compared with the Coulomb gauge fixed field mode spectrum. 
Because $f(| \vec k | ) \propto |\vec A ( \vec k ) |^2$, we obtain `cooled occupation
numbers' from 
\begin{equation}
  f_{\rm cool}(k,\tau) \equiv e^{-2k^2\tau}f(k).
  \label{fcool}
\end{equation}
From this we can calculate the corresponding energy density as
a function of $\tau$.  These are plotted in Fig.~\ref{fg:cooling}
with dashed lines.  We observe that these match the gauge invariant
cooling curves perfectly at initial times where the field amplitudes
are small.

However, at around $t = 28/m_0$ (7th curve from the bottom, see also
Fig.~\ref{fg:strong}) the system enters the non-linear evolution
domain and the two curves start to separate.  This is due to two
effects: firstly, the gauge fixed occupation number calculates energy
slightly incorrectly for large fields, especially in the infrared end
of the spectrum.  Secondly, for large amplitude fields the non-linear
equations of motion make the cooling significantly less efficient in
reducing the energy.  Thus, in the linear approximation used in
Eq.~(\ref{fcool}) the energy decreases much faster
than with the gauge invariant cooling.  This is clearly visible
in Fig.~\ref{fg:cooling}.  Nevertheless, the main
features are the same: the `shoulder' in the cooling curves moves
towards smaller $\tau$, which implies that the ultraviolet modes 
become occupied.

\subsection{Non-weak field initial conditions}
\label{sc:nonWeak}

\begin{figure}[t]

  \vspace*{0.5cm}

  \centerline{
              \epsfxsize=8cm
              \epsffile{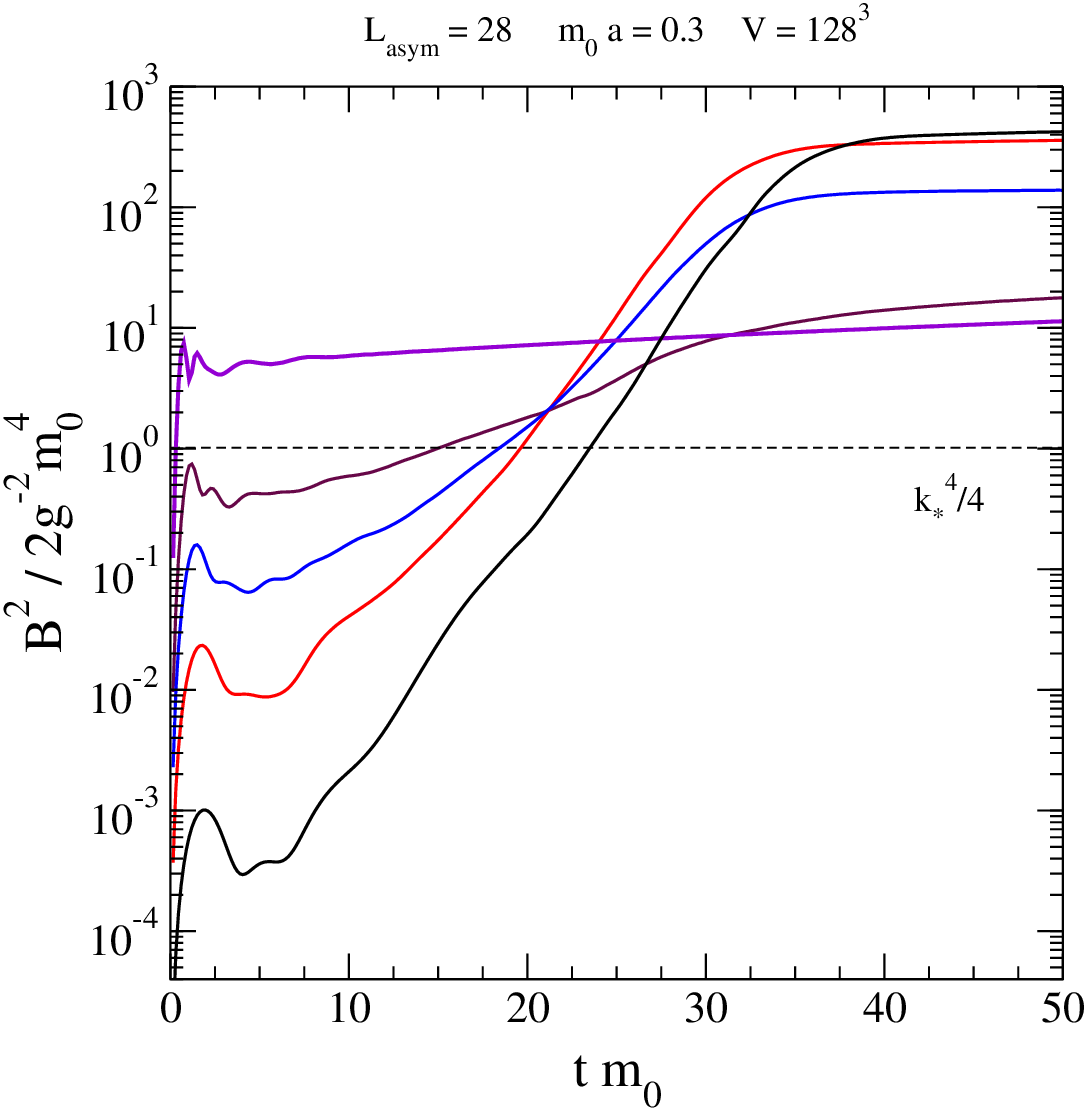}
             }

  \caption{Time evolution of magnetic field energy for different choices of
    the initial field amplitude. 
          }
  \label{fg:largeInitial}
\end{figure}

So far we have only considered very weak initial fields. With
such initial conditions only modes which have $ \vec
k $ very close to the $ z $-axis  get substantially excited because this is where the growth 
rate is the largest.  By the
time the equations of motion become non-linear, the field's momentum 
distribution is almost 1-dimensional. 
It does not mean, however,  
that our results are just what has been 
observed in  $ 1 + 1 $ dimensional simulations \cite{romatschke2d}, where the growth continues
beyond the non-abelian saturation limit. This is  because in our $ 3 + 1 $
dimensional simulations with moderate 
anisotropy the growth saturates even for very weak field initial conditions 
(cf.\ Fig.~\ref{fg:moderate} and Ref.~\cite{arnoldFate}).

Let us now consider larger initial fields
(Fig.~\ref{fg:largeInitial}).  In this case we use only the strong
anisotropy lattices, $\lasym=28$, and $m_0 a=0.3$.  The electric
fields are now initialized with an infrared-dominated spherically
symmetric spectrum, $\langle \vec E ( \vec k ) \rangle \propto \exp[-
\vec k^2/(0.6 m_0)^2]$.  The initial electric field energy densities
vary from $0.0032/(g^{-2}m_0^4)$ to $14.1/(g^{-2}m_0^4)$; from the
electric field the energy is rapidly pumped in the magnetic fields, as
is evident from the figure.  Note that the initial momentum spectrum is
dominated by modes $k\lsim k_\ast$.

We see that there is a very strong dependence on the size of the
initial fields. If the fields start out near the non-abelian point
(\ref{nonabelian}) there is practically no growth.%
\footnote{It should be noted that in this case the system is not
dominated by single mode $\vec k \approx k_\ast \hat\vec z$; thus, the
`non-abelian limit' for energy density does not describe the properties
of the system as well as before.  Nevertheless, we keep this quantity
for comparison.}
This behavior is
very different from the one observed in Ref.~\cite{dumitruAvalanche}
where there is growth for large initial fields.
We leave more detailed analysis for further study.

\section{Lattice artifacts}
\label{sc:lattice}

When a new phenomenon is studied with lattice simulations, it is
very important to quantify possible harmful discretization and
finite volume effects. The very large range of scales makes this check
especially crucial in this case.  As we shall detail below, all 
lattice effects appear to be well under control.

\paragraph{Lattice spacing:~~}
The effects caused by different lattice spacings $a$ were already discussed
above.  As can be seen from Fig.~\ref{fg:moderate}, at weakly anisotropic
hard mode distributions the finite $a$ effects are small -- the 
small dispersion of the results is of the same magnitude than statistical
deviations at fixed $a$.  We made no effort to enforce physically
equivalent initial conditions for different values of $a$.
On the other hand, the finite lattice spacing effects were seen to be 
quite large for strong anisotropy, Fig.~\ref{fg:strong}, due to the
population of the ultraviolet modes.  Even in this case there appears to
be an universal lattice spacing independent evolution, which finite
$a$ simulations follow before they finally saturate.

\begin{figure}[tb]

  \vspace*{0.5cm}

  \centerline{
              \epsfxsize=8cm
              \epsffile{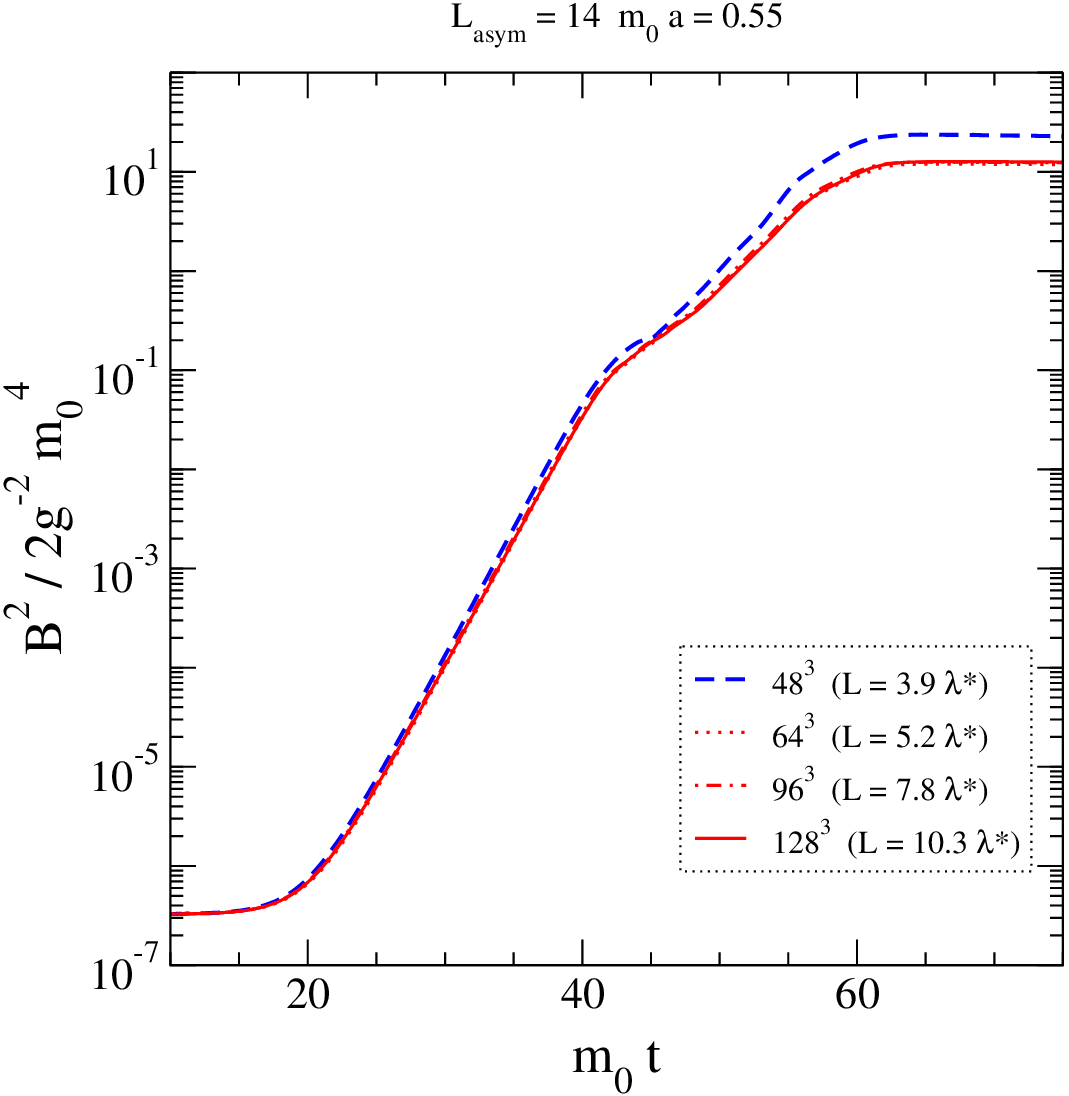}
             }
             \caption{The growth in magnetic energy for 
               $\lasym=14$, $m_0a = 0.55$ runs using different
               volumes.  The 3 largest volume curves are practically
               on top of each other.
             }
  \label{fg:volume}
\end{figure}

\paragraph{Finite volume:~~}
If the volume is too small, it can effectively lower the
dimensionality of the system.  Indeed, too small volume can cause too
much growth.  In Fig.~\ref{fg:volume} we show the evolution using 4
different volumes for $\lasym=14$, $m_0 a = 0.55$ -case.  Except for
the smallest volume the curves fall on top of each other.  (The
statistical dispersion between the large volume runs is very small due
to the smallness of the random initial fluctuations.)  Thus, neither
the exponential growth nor the final saturation can be due to the
finite size of the system.  In general, we require system sizes
$L\gsim 5 (2\pi/k_\ast)$, except for the very smallest lattice
spacing.

\begin{figure}[tb]

  \vspace*{0.5cm}

  \centerline{
              \epsfxsize=8cm
              \epsffile{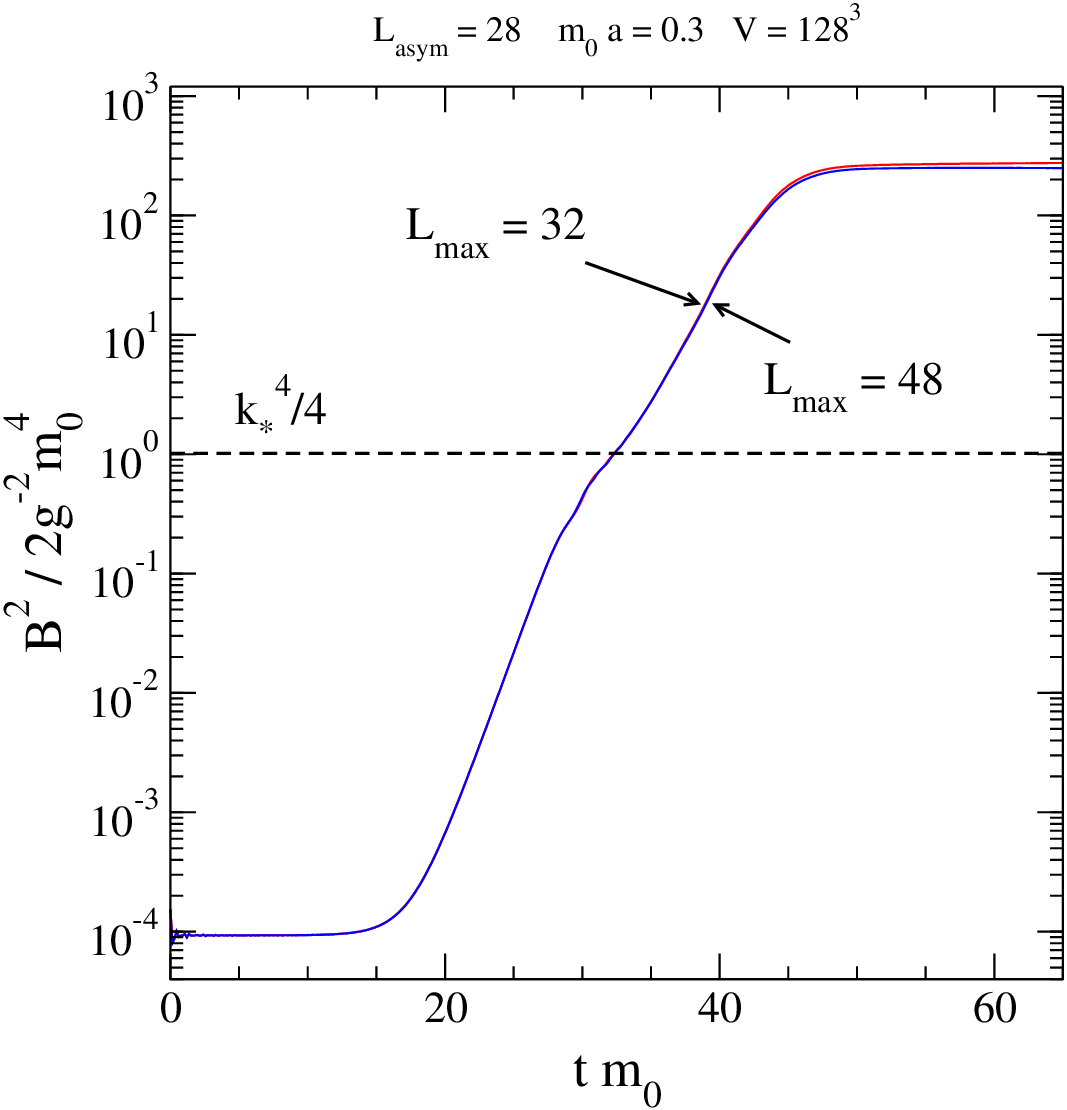}
             }
             \caption{The evolution of the magnetic field energy
               density for $\lasym=28$, $m_0a = 0.3$ -lattices, using
               $\lmax$-cutoffs 32 and 48.  The initial conditions were
               identical for the two runs.}
  \label{fg:lmax}
\end{figure}

\paragraph{Finite $\lmax$:~~}
We have also studied the $ \lmax $-dependence of the field growth. 
In Fig.~\ref{fg:lmax} we show the magnetic energy density evolution for
$ \lasym = 28 $, $ m _ 0 a = 0.3 $ on a $ 128 ^ 3 $ lattice, using
$ \lmax = 32 $ and $ \lmax = 48 $.  In this case we used identical initial
conditions.  As can be seen, the evolution is almost identical.  (See also
Fig.~\ref{rate_L2}.)

\paragraph{Finite $\delta t$:~~}
In addition to finite lattice spacing $a$, in simulations of equations
of motion one has to check the finite update time-step effects.  
In this work we used $\delta t = 0.1 a$, and checked the stability of the
results against $\delta t = 0.05$ simulations with otherwise identical
setup.  The results are in practice indistinguishable, showing that
our original $\delta t = 0.1 a$ is sufficiently small.

\section{Summary and discussion}
\label{sc:summary}

We have studied the dynamics of infrared gauge fields in anisotropic SU(2)
plasmas in the so called hard loop approximation, i.e.,  neglecting the
backreaction of the infrared gauge fields on the phase space distribution of
the high momentum partons. Starting from weak field initial conditions we find
a behavior which appears to be qualitatively different from what was observed
previously for weakly anisotropic plasmas. The field energy grows until 
non-linear effects start playing a role, which slow down the growth. But then
the growth resumes and appears to continue without limit and it is only stopped
by the lattice cutoff. For very strong
anisotropy it is almost as fast as the initial exponential growth. 
This continued growth is different in nature from the linear growth found in
weakly anisotropic plasmas. 

We have studied gauge fixed occupation, gauge invariant operators and
cooling. All methods indicate that there is a rapid transfer of energy to
field modes which have $  |  \vec k  |  \gg k _ { \rm max } $. These are modes
which have no instabilities in the weak field regime. 

For the largest anisotropy we find a growth rate in the strong field
regime which is approximately the same as in the weak field regime.
The growth in total energy persists even though the magnitude of the
soft gauge modes with $|\vec k| \sim k_\ast$ appears to remain
constant.  The mechanism of the energy transfer from the hard modes
($W$-fields) to gauge field modes with $|\vec k| \gg k_\ast$ remains
unknown.

We would like to point out that the earlier 3-dimensional simulations show an
interesting structure which has not been discussed so far. 
After a weak field regime with exponential growth the
system enters a phase where the fields become strong and non-linear effects
become important. But then,  after a brief pause,  the fields again start to  grow
rapidly, almost as fast as during the
initial exponential growth. Only after that there is finally a
saturation and the subsequent linear growth. 
To reiterate, even in the weakly anisotropic case there appears to be a
2-stage structure in the saturation. 
It is conceivable that the
behavior we observed is qualitatively similar. However, we find that 
this continued growth lasts much longer  when we
increase the anisotropy of the system, maybe forever.

{\bf Acknowledgements}
The work of DB was supported in part  through the
DFG funded Graduate School GRK 881.  KR has been partially supported by the
Academy of Finland grants 104382 and 114371.  The simulations in this
work have been made at the Finnish IT Center for Science (CSC, Espoo, Finland).

\end{document}